\documentclass[twoside,twocolumn,9pt]{article}

\usepackage{extsizes}
\usepackage[super,sort&compress,comma]{natbib} 
\usepackage[version=3]{mhchem}
\usepackage[left=1.5cm, right=1.5cm, top=1.785cm, bottom=2.0cm]{geometry}
\usepackage{balance}
\usepackage{mathptmx}
\usepackage{sectsty}
\usepackage{graphicx} 
\usepackage{lastpage}
\usepackage[format=plain,justification=justified,singlelinecheck=false,font={stretch=1.125,small,sf},labelfont=bf,labelsep=space]{caption}
\usepackage{float}
\usepackage{fancyhdr}
\usepackage{fnpos}
\usepackage[english]{babel}
\usepackage{array}
\usepackage{droidsans}

\usepackage{charter}
\usepackage[T1]{fontenc}
\usepackage[usenames,dvipsnames]{xcolor}
\usepackage{setspace}
\usepackage[compact]{titlesec}

\usepackage{epstopdf}
\usepackage{placeins}
\usepackage{multirow}
\usepackage{booktabs}
\usepackage{xr}
\definecolor{cream}{RGB}{222,217,201}

\usepackage{bm}
\usepackage[colorlinks=true,linkcolor=blue,citecolor=blue,urlcolor=blue]{hyperref}

\begin{document}

\pagestyle{fancy}
\thispagestyle{plain}
\fancypagestyle{plain}{
\renewcommand{\headrulewidth}{0pt}
}

\makeFNbottom
\makeatletter

\renewcommand\LARGE{\@setfontsize\LARGE{15pt}{17}}
\renewcommand\Large{\@setfontsize\Large{12pt}{14}}
\renewcommand\large{\@setfontsize\large{10pt}{12}}
\renewcommand\footnotesize{\@setfontsize\footnotesize{7pt}{10}}
\makeatother

\renewcommand{\thefootnote}{\fnsymbol{footnote}}
\renewcommand\footnoterule{\vspace*{1pt}%
\color{cream}\hrule width 3.5in height 0.4pt \color{black}\vspace*{5pt}} 
\setcounter{secnumdepth}{5}

\makeatletter 
\renewcommand\@biblabel[1]{#1}            
\renewcommand\@makefntext[1]%
{\noindent\makebox[0pt][r]{\@thefnmark\,}#1}
\makeatother 
\renewcommand{\figurename}{\small{Fig.}~}
\sectionfont{\sffamily\Large}
\subsectionfont{\normalsize}
\subsubsectionfont{\bf}
\setstretch{1.125} 
\setlength{\skip\footins}{0.8cm}
\setlength{\footnotesep}{0.25cm}
\setlength{\jot}{10pt}
\titlespacing*{\section}{0pt}{4pt}{4pt}
\titlespacing*{\subsection}{0pt}{15pt}{1pt}

\fancyfoot{}
\fancyfoot[RO]{\footnotesize{\sffamily{1--\pageref{LastPage} ~\textbar  \hspace{2pt}\thepage}}}
\fancyfoot[LE]{\footnotesize{\sffamily{\thepage~\textbar\hspace{3.45cm} 1--\pageref{LastPage}}}}
\fancyhead{}
\renewcommand{\headrulewidth}{0pt} 
\renewcommand{\footrulewidth}{0pt}
\setlength{\arrayrulewidth}{1pt}
\setlength{\columnsep}{6.5mm}

\makeatletter 
\newlength{\figrulesep} 
\setlength{\figrulesep}{0.5\textfloatsep} 

\newcommand{\topfigrule}{\vspace*{-1pt}%
\noindent{\color{cream}\rule[-\figrulesep]{\columnwidth}{1.5pt}} }

\newcommand{\botfigrule}{\vspace*{-2pt}%
\noindent{\color{cream}\rule[\figrulesep]{\columnwidth}{1.5pt}} }

\newcommand{\dblfigrule}{\vspace*{-1pt}%
\noindent{\color{cream}\rule[-\figrulesep]{\textwidth}{1.5pt}} }

\makeatother

\twocolumn[
  \begin{@twocolumnfalse}

\LARGE{\textbf{A dual-cutoff machine-learned potential for condensed organic systems obtained \textit{via} uncertainty-guided active learning}}\\
\vspace{0.3cm}\\
\noindent\large{Leonid Kahle,$^{\ast}$\textit{$^{a}$} Benoit Minisini,\textit{$^{a
}$}
Tai Bui,\textit{$^{b}$}
Jeremy T. First,\textit{$^{c}$}
Corneliu Buda,\textit{$^{b}$}
Thomas Goldman,\textit{$^{b}$}
and Erich Wimmer\textit{$^{a}$}} \\
\vspace{0.3cm}\\
\sffamily
\normalsize{Machine-learned potentials (MLPs) trained on \textit{ab initio} data combine the computational efficiency of classical interatomic potentials with the accuracy and generality of the first-principles method used in the creation of the respective training set.
In this work, we implement and train a MLP to obtain an accurate description of the potential energy surface and property predictions for organic compounds, as both single molecules and in the condensed phase.
We devise a dual descriptor, based on the atomic cluster expansion (ACE), that couples an information-rich short-range description with a coarser long-range description that captures weak intermolecular interactions.
We employ uncertainty-guided active learning for the training set generation, creating a dataset  that is comparatively small for the breadth of application and consists of alcohols, alkanes, and an adipate.
Utilizing that MLP, we calculate densities of those systems of varying chain lengths as a function of temperature, obtaining a discrepancy of less than 4\% compared with experiment.
Vibrational frequencies calculated with the MLP have a root mean square error of less than 1~THz compared to DFT.
The heat capacities of condensed systems are within 11\% of experimental findings, which is strong evidence that the dual descriptor provides an accurate framework for the prediction of both short-range intramolecular and long-range intermolecular interactions.
}


 \end{@twocolumnfalse} \vspace{0.6cm}
]

\footnotetext{\textit{$^{a}$~Materials Design SARL, 42 avenue Verdier, 92120 Montrouge, France}}
\footnotetext{\textit{$^{b}$~bp Exploration Operating Co. Ltd, Chertsey Road, 
        Sunbury-on-Thames TW16 7LN, UK}}
\footnotetext{\textit{$^{c}$~bp, Center for High Performance Computing, 225 Westlake Park Blvd, Houston, TX 77079, USA}}

\nocite{rsc-control}
\section{Introduction}
\label{sec:introduction}

Interatomic potentials (IPs) are a long-established method to describe the potential energy and forces\cite{alder_radial_1955} in atomic systems and have provided important insights into the physics of atomic structures throughout the last seven decades.
For organic systems in particular, IPs remained unchallenged for decades because the large system sizes, long time scales, and high configurational entropy made treatment with \textit{ab initio} electronic structure theory impractical.
First developed in the late 1960s and perpetually refined until the present day, classical IPs use well-defined, physics-inspired functional forms to represent the intra- and inter-molecular interactions,~\cite{dauber-osguthorpe_biomolecular_2019,hagler_force_2019} with parameters fitted to reproduce \textit{ab initio} quantum mechanical and experimental data. 
This road of continuous improvement led to the creation of the well-known EAM,~\cite{daw_embedded-atom_1984} AMBER,~\cite{case_ambertools_2023} OPLS,~\cite{jorgensen_potential_2005} CHARMM,~\cite{brooks_charmm_1983} and CFF~\cite{maple_derivation_1994} force fields.
Nevertheless, the rigorous functional form of these force fields limits the generality of each.~\cite{hagler_force_2019}
Moreover, most classical IPs are non-reactive and constrained to certain classes of materials, limiting the usability in applications that combine different classes of materials.\cite{di_pierro_optimizing_2015}
As an example, calculating the interface properties between organic liquids and solid metals with IPs remains challenging because each material is best described by a different class of force field.~\cite{dasetty_simulations_2019}

Machine-learned  potentials (MLPs) promise to overcome these limitations.~\cite{behler_representing_2007, behler_generalized_2007, behler_perspective_2016,fiedler_deep_2022}
In general -- and inspired by the applications of machine learning in many other domains such as image recognition or language models -- MLPs use non-linear regression techniques such as neural networks (NNs) to fit the potential energy surface (PES) of a system, using only atomic positions and species as input to the framework.~\cite{eyert_machine-learned_2023}
The total energy of the system is decomposed into atomic contributions, which again are obtained as a function of the atomic environment represented by a descriptor:
\begin{equation}
    E = \sum_i E_i = \sum_i f(\mathbf{R}_i, \sigma_i, \{\mathbf{R}_j, \sigma_j\}; \theta),
\end{equation}
where $\mathbf{R}$ and $\sigma$ are the positions and species, respectively, of atom $i$ and of all nearby atoms $j$, as defined by a cutoff radius $r_c$, and $f$ is the function described by the MLP parametrized by a set of parameters $\theta$.
The regression to obtain the optimal set of parameters $\theta$ that define $f$ is usually performed on datasets containing energies, forces, and potentially stresses calculated from density-functional theory (DFT)~\cite{hohenberg_inhomogeneous_1964,kohn_self-consistent_1965} or other first-principles methods.
To be useful, a MLP needs to reproduce the accuracy and generality of \textit{ab initio} quantum mechanical computations at the computational efficiency of classical force fields. In particular, they need to scale linearly in compute time with system size.~\cite{zuo_performance_2020}
MLPs should be -- by design -- symmetric with respect to translation, rotation, and atomic permutation of the system; in other words, they should possess SO(3) invariance.
Almost all MLPs used in recent studies possess this invariance as it allows for greater data sparcity, among other benefits.~\cite{ceriotti_unsupervised_2019}
Differences in MLP architectures are mostly due to the application of different descriptors of the atomic environment and/or the use of different regression techniques.
Examples include the atom-centered symmetry functions that form a SO(3) invariant input to a NN architecture,~\cite{behler_atom-centered_2011,behler_representing_2014} the Coulomb matrix,~\cite{hansen_assessment_2013,bereau_transferable_2015}
the Smooth Overlap of Atomic Position (SOAP) descriptor that usually serves as input to a Gaussian process regression leading to the Gaussian Approximation Potential (GAP),~\cite{bartok_gaussian_2010,rowe_development_2018,thiemann_machine_2020}
the Spectral Neighbor Analysis Potential (SNAP) that is commonly linearly regressed,~\cite{thompson_spectral_2015,wood_extending_2018} and emerging graph-convolution neural network architectures.~\cite{schutt_schnet_2018,thomas_tensor_2018,xie_crystal_2018,batzner_e3-equivariant_2022,musaelian_learning_2023}
Additionally, the atomic-cluster expansion (ACE), developed and implemented by Drautz \textit{et al.}, provides an extendable general descriptor to arbitrary order~\cite{drautz_atomic_2019,lysogorskiy_performant_2021,bochkarev_efficient_2022} that can be used in linear models, simple non-linear models, or in message-passing frameworks.~\cite{kovacs_evaluation_2023}
In summary, a plethora of MLP architectures provide rigorous and easily automated frameworks to fit the PES of atomic systems, given enough data.

An important challenge to designing accurate and robust MLPs can be the large number of possible atomic configurations in many applications, resulting in a complicated and rugged function to learn.
While this is manifested in all materials science domains, the high entropy of condensed organic systems makes this a particular challenge.
In such cases, the flexibility of MLPs is also their curse, as it can lead to poor extrapolative behavior,~\cite{behler_first_2017} resulting in low accuracy in configurations that are outside the training-set distribution, as well as instabilities during molecular dynamics (MD) simulations.
Recently, the stability of MLP-driven MD simulations has drawn attention as an important but undervalued metric in MLP design.~\cite{fu_forces_2022}
Indeed, a study by Stocker \textit{et al.}~\cite{stocker_how_2022} on organic systems showed that while a relatively low number of training structures may be sufficient to accurately describe the energies and forces of an independent test set, achieving stability in an MD simulation required orders of magnitude more training data;
in their study, based on the QM7-x training set, the \textit{GemNet} MLP required 3,200 training configurations to generate a low-error potential, but 320,000 training configurations to achieve stability in an MD simulation.

The flexible nature of MLPs requires ensuring that the MLP is accurate for all possible configurations that are visited in simulation. 
To tackle that challenge, an important hurdle to overcome is whether an MLP can predict if it is -- or should be -- uncertain in its prediction of energy and forces for a given configuration.
In this work, we use uncertainty to refer to \textit{epistemic} uncertainty, due to a lack of data, as opposed to \textit{aleatoric} uncertainty, inherent to the data itself.~\cite{yarin_gal_uncertainty_2016,vandermause_--fly_2020}
Uncertainty can be calculated in different ways depending on the MLP employed: Gaussian process models (\textit{e.g.}, GAP) provide an uncertainty inherently, whereas neural network models use committee or ensembles deviation~\cite{jeong_efficient_2020, kahle_quality_2022,schran_committee_2020,zhang_dp-gen_2020,busk_calibrated_2021,schran_machine_2021,kahle_quality_2022} or dropout techniques~\cite{wen_uncertainty_2020} for an uncertainty estimate.

The uncertainty value estimated by a model can be either used to switch to a more general computational method or to inform on missing configurations in the training set.
The former, commonly referred to as learn on-the-fly (LOTF) techniques, uses an estimate of model uncertainty to switch between data-trained models (such as MLPs) and reliable DFT predictions during MD simulations.~\cite{li_molecular_2015,jinnouchi_fly_2019, jinnouchi_making_2020,vandermause_fly_2020}
In contrast to LOTF, Active Learning (AL) techniques allow for the steered selection of new configurations after exploration.~\cite{schran_machine_2021}
The selection process may be guided by quantifying the model uncertainty and algorithmically selecting and labeling the most informative data points from an unlabeled dataset, thereby significantly reducing the training set size.~\cite{csanyi_learn_2004}

AL for MLPs usually consists of a sequence of three steps that run in a loop until convergence:
\begin{enumerate}
    \item \textit{Exploring}: An MLP is used to explore new configurations (\textit{e.g.}, \textit{via} MD at different temperatures and pressures) and to quantify the uncertainty of the prediction for the sampled configurations.
    \item \textit{Labeling}: Structures with new information -- as determined by a high uncertainty estimated in the previous step -- are  labeled with an \textit{ab initio} method and added to the training set. 
    \item \textit{Training}: A new model is trained with the augmented training set.
\end{enumerate}
The iterative process can improve the model performance in a data efficient manner, because, by definition, only useful training data are added at every iteration.
For tasks where labeling data is costly, such as labeling with electronic-structure methods, the reduction of training set size has clear advantages.

Another challenge when designing MLPs for organic systems is the presence of long-range van-der-Waals (vdW) and electrostatic interactions.
The treatment of long-range interactions with an MLP typically falls into one of three categories:~\cite{anstine_machine_2023, behler_four_2021} (1) the long-range interactions are neglected, (2) the long-range interactions are described by parameters depending only on the local environment, or (3) the long-range interactions are incorporated from the global environment.
While many publications in the literature show good agreement with experimental data using MLPs of type (1), even in some systems where long-range contributions should be significant,\cite{marcolongo_simulating_2019} there is consensus that the inclusion of long-range interactions improves the accuracy and robustness of a MLP.~\cite{artrith_high-dimensional_2011,anstine_machine_2023, behler_four_2021,yue_when_2021}
Still, due to the complications of designing MLPs of type (2) and (3), applications of MLPs to organic systems have often disregarded the treatment of condensed phases and focused on properties where intermolecular interactions can be neglected; this is underscored by many excellent studies that report high accuracy of MLPs in the treatment of single organic molecules in vacuum.~\cite{chmiela_towards_2018, batzner_e3-equivariant_2022}
In cases where the description of intermolecular interactions is qualitatively accurate, condensed phases can be modeled in constrained-volume simulations.
Indeed, MLPs developed for different molecules such as water, acetonitrile, and methanol\cite{maldonado_modeling_2023} or a mixture of choline chloride and urea\cite{shayestehpour_efficient_2023} have found success in reproducing condensed phase properties such as the radial-distribution function (RDF).
In addition, Zaverkin \textit{et al.} showed that transfer learning increased the robustness of MLPs for constant-volume, constant-temperature (NVT) simulations of deca-alanine in water.~\cite{zaverkin_transfer_2023}
Nevertheless, effects of temperature and pressure on molecular densities cannot be estimated from constrained-volume simulations, and the  trustworthiness of the intermolecular interactions in such cases should be considered.

Few publications attempt to describe and study properties emergent from weak long-range interactions (\textit{i.e.}, vdW and electrostatic), such as the calculated density from MD simulations in the isothermal-isobaric (NPT) ensemble.  
Morado \textit{et al.} provide an in-depth comparison of the advantages and disadvantages of MLPs for $\gamma$-fluorohydrins~\cite{morado_does_2023} against a classical force field, concluding that the MLP described small molecules well but could not capture dispersion interactions.
A more successful example was provided by Magd\u{a}u \textit{et al.},~\cite{magdau_machine_2023} who developed a model based on the GAP, which generated stable dynamics in the NPT ensemble for ethylene carbonate (EC) and ethyl-methyl carbonate (EMC) binary liquid models.
By combining a short-range kernel with a simpler, long-range kernel (both relying on GAP but capturing different length scales), the authors obtained an overestimation of 5.5\% and an underestimation of 2\% on the density for pure EC and EMC, respectively.
Interestingly, Abedi \textit{et al.} accurately predicted the liquid vapor coexistence curve for methane with a high dimensional NN potential without adjusting the architecture by using a relatively large cutoff of 6.25~\AA.\cite{abedi_high-dimensional_2023}
Hajibabaei \textit{et al.}~\cite{hajibabaei_machine_2021} used LOTF to prepare a training set of different alkanes and alkenes,
and the resulting MLP accurately reproduced the heat of vaporization of ethane.
As an additional example, the developers of MACE-OFF23~\cite{kovacs_mace-off23_2023}  trained a transferable MLP for organic molecules that accurately predicted densities and the heats of vaporization for a benchmark set of 109 molecules.
We suspect that the reason MACE-OFF23 captured long-range interactions so well was due to the use of a message-passing architecture, which effectively multiplies the interaction or cutoff radius of the MLP by the number of message-passing layers.
Finally and most recently, Gong \textit{et al.}~\cite{gong_bamboo_2024} implemented and trained the BAMBOO framework, which achieved high accuracy in reproducing the densities of organic battery electrolytes by using a two-fold approach: including an explicit vdW function into their MLP model, and including experimental densities into the MLP fitting approach.
In summary, recent work shows that most standard approaches for MLP training are inadequate for the treatment of systems with vdW interactions, requiring a significant increase in the cutoff radius (or effective cutoff radius in message-passing frameworks), multiple-descriptor architectures, and a high number of training configurations.

In this work, we derive and showcase a novel approach based on the ACE descriptor to producing a 
data-efficient MLP for condensed organic systems using an uncertainty metric for configuration selection.
Following the work of Magd\u{a}u \textit{et al.} using GAPs,~\cite{magdau_machine_2023} we describe a protocol for combining a rich, short-range descriptor with a more basic, long-range descriptor, all within an ACE-MLP.
In addition, we employ uncertainty-driven AL to guide the exploration towards configurations that provide new, useful information to the MLP.
We use the MLP to calculate phonon-frequencies and run MD simulations in the NPT ensemble to obtain properties such as the densities of the liquid phase, RDFs, heats of vaporization, and heat capacities.
We compare our results to experimental findings, providing evidence that a dual-cutoff MLP architecture, coupled to uncertainty guided AL for the training set generation, is well-suited for complicated systems such as condensed phase organic systems, where both short- and long-range contributions to the energies and forces are important.
This study opens avenues for many such complex systems to be modeled with MLPs in a data-efficient manner that may be otherwise difficult to study with empirical force fields.

\section{Methods}

\begin{figure}[t!]
    \centering
    \includegraphics[width=\hsize]{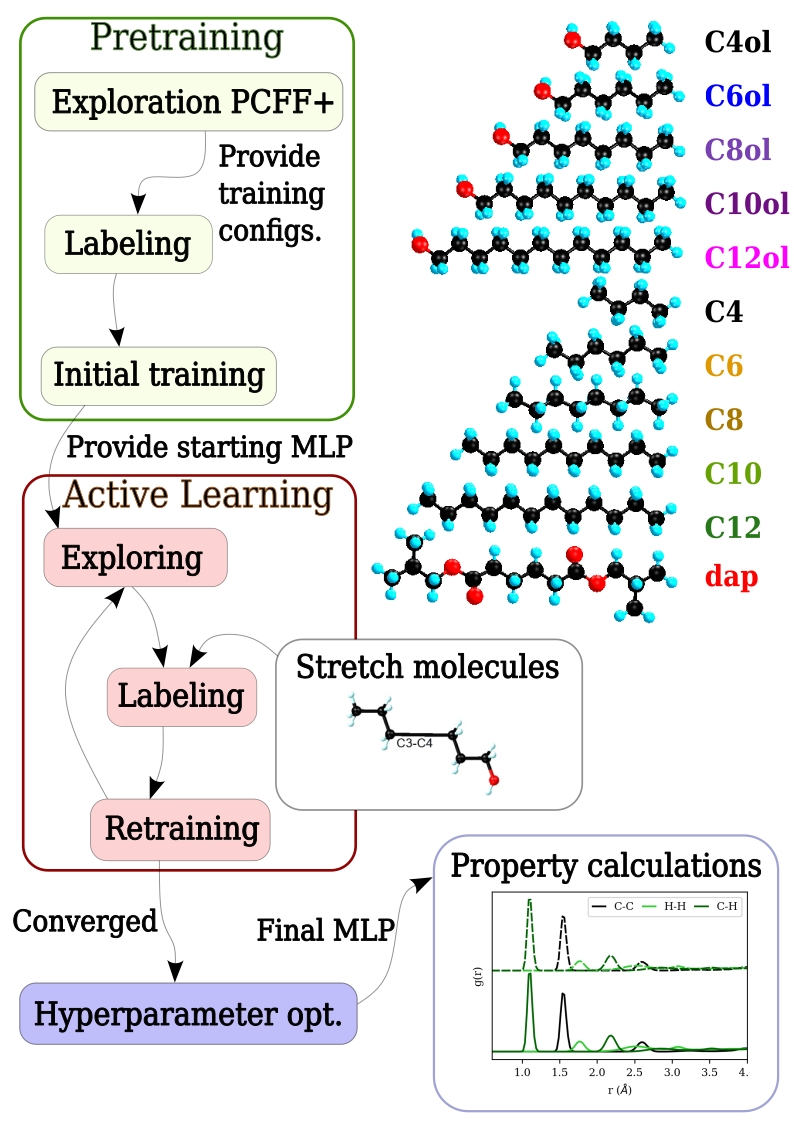}
    \caption{The overall workflow employed in this work. Green box: Initial training configurations are generated with the PCFF+ force field, labeled with DFT, and used to train a first MLP.
    Red box: The MLP is then used to explore new configurations, which are labeled with DFT, added to the training set, and the model retrained. 
    Grey box: Additional configurations were explicitly added to explore the fragmentation PES, as described in \autoref{sec.methods.active}. 
    After the \textit{Active Learning} cycle converged, the hyperparameters of the MLP were optimized (blue-filled box), and the best model was evaluated on a variety of properties (blue-outlined box).
    In the top right, we show single molecule representations of the systems of interest.
    From top to bottom: butanol, hexanol, ocatanol, decanol, dodecanol, butane, hexane,  octane, decane, dodecane, and diisobutyl adipate.
    Black spheres represent carbon atoms, red spheres represent oxygen atoms, and blue spheres represent hydrogen atoms.
    Each molecule is labeled with its abbreviation used throughout this manuscript, and the color of the label corresponds to the color of all corresponding data in the subsequent figures, where applicable.}
    \label{fig:methods.workflow}
\end{figure}

The workflow employed in the present investigation
is illustrated in \autoref{fig:methods.workflow}.
For the systems of interest (\autoref{sec.methods.systems}), we generated initial training configurations with the PCFF+ force field,~\cite{rigby_computational_2016} presented in \autoref{sec.methods.forcefield}.
These training configurations were labeled using an accurate DFT protocol (\autoref{sec.methods.DFT}) and used to train a first MLP with the ACE descriptor (\autoref{sec.methods.ace}).
We chose the ACE descriptor because a) recently published benchmarks show good accuracy for organic molecules,~\cite{batzner_e3-equivariant_2022} b) a proven uncertainty metric is implemented, and c) the MLP training is implemented in the \texttt{pacemaker} code~\cite{lysogorskiy_performant_2021, bochkarev_efficient_2022} with the option of setting different cutoffs for different two-body terms.
This MLP was used to start the AL cycle (\autoref{sec.methods.active}).
After the AL cycle converged, the hyperparameters of the model were optimized (\autoref{sec.methods.hyperparameters}), and the best model was evaluated on a variety of properties (\autoref{sec.methods.analysis}).

\subsection{Systems under investigation}
\label{sec.methods.systems}

In this work, we focused on the development of a MLP for condensed organic systems. We targeted linear alkanes and alkanols as representative of the simplest building blocks of many organic systems. In addition, a more complex system, diisobutyl adipate, was also included to test the generalizability of the model. Namely, our systems of interest are butanol (C4ol), hexanol (C6ol), octanol (C8ol), decanol (C10ol), butane (C4), hexane (C6), decane (C10), and diisobutyl adipate (dap), as illustrated in \autoref{fig:methods.workflow}. 
To train the MLP for condensed systems, a periodic box size of (10~\AA)$^3$ was chosen because it was amenable to DFT calculations, allowed for chain-chain interactions, but limited molecular self-interaction. 
This box size was used for all systems, resulting in different numbers of molecules for each system type. We refer to these systems with a -$n$ suffix, where $n$ indicates the number of molecules in the (10~\AA)$^3$ box. 

\subsection{Initial training configurations}
\label{sec.methods.forcefield}
We first generated training configurations of C4ol-8, C6ol-6, C8ol-4, C10ol-3, and dap-3 \textit{via} MD in the NPT ensemble using the PCFF+ force field at different temperatures (from 298~K to 373~K) and pressures (1~atm, 10~atm, and 100~atm). 
PCFF+ was chosen to allow the generation of diverse snapshots in a computationally efficient manner, compared to running \textit{ab initio} MD.
Configurations were selected by sampling every 10~ps, resulting in 200 configurations of C4ol-8, 50 configurations of C6ol-6, 50 configurations of C8ol-4, 50 configurations of C10ol-3, and 330 configurations of dap-3.
In order to increase the configuration space represented by the training set, we augmented each sample set with additional configurations generated by adding random displacements (sampled from the normal distribution with a 0.05~\AA~standard deviation) to the MD generated configurations. 
50 such configurations were added for all alkanols and alkanes, and 110 configurations were added for dap-3.
Finally, in order to teach the model the difference between isolated molecules and condensed phases (\textit{i.e.}, intermolecular interactions), 50 configurations for C4ol-1 and 90 configurations for dap-1 (\textit{i.e.}, single molecules) were generated from NVT simulations (driven by PCFF+) in a  periodic box, sampling every 10~ps, and without adding random displacements.
The size of the box was determined by the minimum and maximum positions of the atoms in the configuration, adding $8$~\AA~to each dimension to ensure enough vacuum was present.
All configurations selected for the training set were labeled \textit{via} our DFT protocol presented in \autoref{sec.methods.DFT}.

\subsection{Density functional theory calculations}
\label{sec.methods.DFT}

For all DFT calculations, we used the \texttt{VASP} code~\cite{kresse_efficient_1996, kresse_efficiency_1996} with the precision set to ``Accurate'', a plane  wave cut-off at 500~eV.
For all structures in the training set, the $\Gamma$-point alone suffices to sample the Brillouin zone when labeling training configurations at lattice constants of 10~\AA ~or larger, speeding up calculations because it allows for real-value numerical operations.
All DFT calculations were performed within the \textit{MedeA} software environment.~\cite{eyert_unravelling_2018}

DFT calculations of consistent accuracy and precision are required to provide data to train the MLP and to validate it.
The prevalence of intermolecular interactions in condensed organic systems required the inclusion of dispersive interactions in the training set.~\cite{liu_how_2012}

To understand which DFT functional was best suited for providing high-accuracy training data, we compared the ground-state cell parameters with experimental data for a variety of compounds with similar chemical moieties as in our systems of interest, taking inspiration from previous work that used peptides and organic crystals to assess the accuracy of force fields~\cite{hagler_force_2019} and vdW functionals.~\cite{pham_density-functional_2016, liu_how_2012}
We first compared the cell parameters of crystalline polyethylene and methanol obtained by different DFT exchange-correlation functionals  with experimental data obtained at 4~K~\cite{avitabile_low_1975} and 122~K.~\cite{kirchner_cocrystallization_2008}
The exchange-correlation functionals presenting the lowest deviation between calculated and experimental equilibrium lattice parameters were selected, and the equilibrium lattice parameters of isopropanol,~\cite{ridout_low-temperature_2014}
n-octanol,~\cite{shallard-brown_n-octanol_2005}
ethanol,~\cite{jonsson_hydrogen_1976}
3-ethyl-3-pentanol,~\cite{bond_3-ethylpentan-3-ol_2006}
butanol,~\cite{derollez_structure_2013}
and decane~\cite{bond_n-decane_2002} were evaluated to validate our choice of functional.
Based on these results, we used revPBE-vdw\cite{dion_van_2004,roman-perez_efficient_2009,klimes_van_2011} for all subsequent DFT calculations, including the labeling of the training configurations. Additional details and results of the benchmark calculations are provided in Section S.1 of the Supplementary Information.

Our finding that the revPBE-vdw functional reproduces the dispersion interaction between alkane chains in organic systems is consistent with recent publications.
For example, Tsuzuki \textit{et al.}~\cite{tsuzuki_accuracy_2023} showed that the intermolecular interaction energies calculated on three hydrocarbon complexes deviate by less than 10\% in comparison to CCSD(T) results.
While Klime\v{s} \textit{et al.}~\cite{klimes_chemical_2009} showed that revPBE-vdw poorly balanced H-bonding and dispersion interactions in the $\pi$-stacked molecules present in the S22 dataset,~\cite{jurecka_benchmark_2006} we note that such interactions are not germane to our systems of interest.

\subsection{The Atomic Cluster Expansion (ACE) descriptor}
\label{sec.methods.ace}

The ACE descriptor was envisioned by Drautz~\cite{drautz_atomic_2019} and implemented by Lysogorskiy \textit{et al.}~\cite{lysogorskiy_performant_2021, bochkarev_efficient_2022} into the \texttt{pacemaker} code.
Below, the ACE descriptor is briefly summarized before presenting the dual-particle ACE descriptor implemented in this work.

A one particle basis function $\phi$ is built up from a set of radial and spherical functions of the distance vector, $\bm r_{ji}$, between two atoms $i$ and $j$, along with its norm $|\bm r_{ji}|$ and unit vector $\hat{\bm r}_{ji} = \frac{\bm r_{ji}}{|\bm r_{ji}|}$:
\begin{equation}
    \phi_{\mu_i \mu_j n l m} = R^{\mu_i \mu_j}_{nl} ( |\bm r_{ji} | ) Y_{lm} (\hat{\bm r}_{ji}),
    \label{eq.one-particle-basis}
\end{equation}
where $\mu_i$ indexes the species of atom $i$, and $n$, $l$, and $m$ are the radial, angular, and spin numbers of the spherical-harmonics basis functions.
The radial basis can be selected from different basis functions,~\cite{bochkarev_efficient_2022} the default being spherical Bessel functions which we employ in this work.
These basis functions are attenuated with a cutoff function that smoothly reaches a value of 0 at $|\bm r_{ji}| = r_c$ in the range $r_c - \Delta < r < r_c$. This polynomial cutoff function is: $\frac{1}{2}\left(1 + ax + bx^3 +c x^5\right);$ with $x=1-2\left(1+ \frac{r-r_c}{\Delta}\right)$ and $a=1.875$, $b=-1.25$, $c=0.375$, which ensures gradients of 0 at $r=r_c - \Delta$ and $r=r_c$.

An atomic basis function is built for every atom by taking the sum over all atoms $j$ within the cutoff $r_c$:
\begin{equation}
    A_{i\mu n l m} = \sum_j \delta_{\mu \mu_j}  \phi_{\mu_i \mu_j n l m}.
\end{equation}
The vector $\bm A$ is invariant to translation because only interatomic distances are used, and invariant to permutation because of the summation over all neighbors, but not yet invariant to rotation. The latter invariance is achieved by calculating a many-body basis function $\bm B$ as a product over components of $\bm A$ up to order $\nu$:
\begin{equation}
    B_{i\mu_{1\dots\nu} n_{1\dots\nu} l_{1\dots\nu}} = C(l_{1\dots\nu}, m_{1\dots\nu}) \prod_{t=1}^{\nu} A_{i\mu_t n_t l_t m_t},
    \label{eq.def-B}
\end{equation}
where the index $x_{1\dots\nu}$ specifies that a tuple of length $\nu$ of indices is used (\textit{e.g.}, $x_1x_2x_3$ for $\nu=3$). $C(l_{1\dots\nu}, m_{1\dots\nu})$ refers to the generalized Clebsch–Gordan coefficients that ensure that the summation of indices $m$ and $l$ results in rotation-invariant expansion, making $B$ also rotationally invariant.
The atomic property vector $\varphi_i^{(p)}$ is constructed from $\bm B$ as:
\begin{equation}
    \varphi_i^{(p)} = \sum_{\mu_{1\dots\nu} n_{1\dots\nu} l_{1\dots\nu}} c_{\mu_{1\dots\nu} n_{1\dots\nu} l_{1\dots\nu}}^{(p)} B_{i \mu_{1\dots\nu} n_{1\dots\nu} l_{1\dots\nu}},
\end{equation}
where $\bm c^{(p)}$ is a vector of free model parameters that need to be regressed to minimize the loss function.
Finally, the atomic energy $E_i$ is described as a Finnis-Sinclair-type expansion of two atomic property vectors:
\begin{equation}
    E_i = \varphi_i^{(1)} + \sqrt{\varphi_i^{(2)}}.
    \label{eq.energy-ACE}
\end{equation}

The potential energy of the system is obtained as the sum over all atomic energies in the system $E_{pot} = \sum_i E_i$, and the force acting on atom $i$ is taken as the derivative of the potential energy: $\bm F_i= -\frac{\partial E_{pot}}{\partial \bm R_i}$. Finally a loss function is defined for the regression:
\begin{align}
    \notag \mathcal{L} =& (1-\kappa) \left(\sum_n \left(\frac{E_{pot, ACE}^{(n)} - E_{pot, label}^{(n)}}{N_{at}^{(n)}}\right)^2 \right) \\
    \notag
    &+ \kappa \sum_n \sum_i^{N_{at}^{(n)}} (|\bm F_{i, ACE}^{(n)} - \bm F_{i, label}^{(n)}|^2) \\
    &+ \lambda \sum_p \sum_{\mu_{1\dots\nu} n_{1\dots\nu} l_{1\dots\nu}} \left(c_{\mu_{1\dots\nu} n_{1\dots\nu}  l_{1\dots\nu}}^{(p)}\right)^2,
    \label{eq.loss}
\end{align}
where $n$ iterates over configurations in the training set, $i$ iterates over atoms in a configuration, $\kappa$ is the weight of the forces, and $\lambda$ is the weight of the $l_2$ regularization.

In order to create a descriptor that captures both short- and long-range interactions, we built an ACE descriptor with two separate cutoffs for the pair terms in Eq.~\eqref{eq.one-particle-basis}:
a short-range cutoff $r_c^{s}$ and a long-range cutoff $r_c^{l}$.
These cutoffs are hyperparameters of the MLP, were optimized (\autoref{sec.methods.hyperparameters}), and were found to be 3~\AA ~and 7~\AA, respectively, in the best performing MLP.
The long-range cutoff was only applied to one specific pair term, namely the ``C-C'' interaction term between carbon atoms, and is carried into the descriptor (see Eqs.~\eqref{eq.one-particle-basis} to \eqref{eq.energy-ACE}).
As a result, a ``dual'' descriptor is obtained that includes both information about all atomic interactions at short-range and limited information of neighboring carbon atoms at long-range. 
The nature of the cluster expansion in ACE mixes the long-range information into higher-order three-body (and higher) terms.

\subsection{Active Learning}
\label{sec.methods.active}

\subsubsection{Initial training set}
The AL process was seeded by a training set of C4ol-8 and C4ol-1 configurations, as generated in \autoref{sec.methods.forcefield}. 
Next, a MLP was iteratively trained by cycling \textit{exploring} -- \textit{labeling} -- \textit{training} steps, as described below and illustrated in \autoref{fig:methods.workflow}.
Simultaneously, the complexity of the systems was increased by adding in PCFF+ generated configurations of C6ol-6, C8ol-4, C10ol-3, dap-1, and dap-3.
The training sets at each stage of the AL process are summarized in Table S1 of the Supplementary Information. 
The underlined configurations were generated with PCFF+, the configurations in bold  were created during the exploration phase of the MLP, and the configurations in italics were created by the \textit{fragmentation} algorithm outlined is \autoref{sec.methods.active.exploring.fragmentation}.

\subsubsection{Exploring} 
\label{sec.methods.active.exploring}
We used the extrapolation grade $\gamma$ defined by Lysogorskiy \textit{et al.}~\cite{lysogorskiy_active_2023} as our uncertainty metric.
The extrapolation grade is a per-atom property that defines whether an atomic configuration, given by its descriptor $B_{i\mu_{1\dots\nu} n_{1\dots\nu} l_{1\dots\nu}}$ (see Eq.~\eqref{eq.def-B}), can be represented by the active set already in the training set.
If $\gamma > 1$, the atomic environment is not contained in the volume spanned by the active set, which we interpret as the extrapolative regime.
The extrapolative grade $\gamma$ is an per-atom property, so we define $\gamma_{max}$ as the largest $\gamma$ in a configuration, which is used to determine the configurations for labeling.


New training configurations were explored using the MLP model from the previous iteration.
For each \textit{exploring} phase, we sampled from NPT ensembles generated at  1~atm and 273~K, 313~K, and 393~K.
To force exploration of smaller bond distances and intermolecular distances, we additionally sampled from an ensemble generated at 100~atm and 273~K.
The length of the simulation and the employed sampling frequency were adaptively selected, always attempting to sample 100 snapshots at a given sampling frequency, starting at 160~fs.
If the MLP produced robust dynamics, the sampling frequency was halfed and the total simulation time doubled.
On the contrary, if the MLP did not produce robust dynamics, the sampling frequency was doubled and the total simulation time was halfed.
Our definition of robustness and the thresholds employed in this work are presented in Section~S3 of the Supplementary Information 
for the interested reader.
A maximum sampling frequency was set to 1~ps and if no extrapolative structures were found, the MLP was labeled as sufficiently trained for that structure under those thermodynamic conditions.
The numbers of configurations of each system added to the training set during the \textit{exploring} phases are tabulated in Table S1 of the Supplementary Information 
for each iteration.

\paragraph{Bond dissociation}
\label{sec.methods.active.exploring.bondbreaking}
During our training cycle, we perceived instabilities in the MD simulations for our \textit{exploring} phases, especially with large systems, despite observing good accuracy in energies and forces of test configurations. 
This is consistent with Stocker \textit{et al.},~\cite{stocker_how_2022} who observed unstable dynamics in otherwise accurate potentials.
Therefore, complementary to the uncertainty-guided exploration, we implemented an algorithm to actively detect unphysical bond lengths in the MD simulations by tracking interatomic distances of atoms that are conceptually bonded.
A bond was considered unphysical if the bond distance was below 0.6~\AA\ or above 2.6~\AA, regardless of the involved species.
Any configurations with unphysical bond lengths were added to the training set in order to increase the robustness of the MLP.

\paragraph{Fragmentation}
\label{sec.methods.active.exploring.fragmentation}
Even after the inclusion of bond-breaking events, we still observed instabilities in MD simulations with the intermediate ACE-MLPs: while the ACE-MLPs at small system sizes (\textit{i.e.}, roughly (10~\AA)$^3$ and amenable to DFT) could stably reach 100~ps in the trainings, larger systems or longer simulations could be unstable.
The unphysical bond breaking at ambient conditions was puzzling, since the MLP was fit to configurations of the ground state and included configurations with long bond lengths and relatively high energy.
This observation could be explained by the enlarged space of possible configurations, as well as the higher probability of excitations that are possible in larger systems, taking an inference configuration out of the safe envelope of training configurations (where the MLP is interpolating rather than extrapolating).
Investigating this further, we found that the ``fragmentation energy'' (\textit{i.e.}, the energy profile along a specific bond distance) remained inaccurate and the energy of dissociation was underpredicted. 
We reasoned this was likely due to a lack of training configurations that contain such fragments, and 
in order to more intentionally explore the fragmentation PES, we explicitly added ``fragments'' to the training set in AL cycles 11, 12, and 17. 
Fragments were generated by isolating molecules from the condensed-phase configurations in our training set, randomly selecting a bond, and then stretching the bond by a factor between 1 to 3 (chosen from a uniform random distribution). The inclusion of ``fragments'' resolved the underprediction of the fragmentation energies, as discussed further in Section S.2 
and demonstrated in Fig. S4 in the Supplementary Information.

\subsubsection{Labeling} 
All selected configurations from the \textit{exploring} phase were labeled with the DFT parameters outlined in \autoref{sec.methods.DFT}.

\subsubsection{Training}
The parameters of the MLP may either be initialized from a random distribution or from the parameters regressed in the previous iteration of AL, if existing. 
We chose the latter option, which reduced compute time since the model parameters were already fit to a similar data set and only needed, in most cases, slight adjustments.
Training was performed with the \texttt{pacemaker} code, and a typical input file containing all parameters is presented in Section S.4 of the Supplementary Information.
Of note, we used a low $\kappa$ of 0.001 (further discussed in \autoref{sec.methods.hyperparameters}) and only 500 training iterations each cycle, since the starting parameters were initialized from the previous AL cycle.


\subsubsection{Training set refinement}
As the final step to the training set generation, we removed all configurations from the training set that had any atomic force with a norm above 10~eV\,\AA$^{-1}$, since we observed deterioration of the MLP accuracy when distorted structures with larger force components were included.

\subsubsection{Hyperparmater optimization}
\label{sec.methods.hyperparameters}
After completing the training set generation, we performed a hyperparameter optimization to optimize the MLP hyperparameters given all training configurations.
We varied the number of $n$ and $l$ components, the total number of parameters, the short and long cutoff, the weight of forces in the loss function $\kappa$, and the  weight of the l2-regularization $\lambda$ of the model (see Eq.~\eqref{eq.loss}).
In addition, we tested whether long-range contributions should be described by hydrogen-hydrogen or carbon-carbon pair interactions.
We selected the MLP that produced the most accurate energies and forces on a test set, and this MLP was analyzed throughout the rest of this work. 
The energies and forces predicted for the test set (2\% of the total configurations) are given in Fig. S6 of the Supplementary Information. 
The RMSE estimated on the test set was 3.22~meV/atom for the potential energy and 95.65~meV\,\AA$^{-1}$ on atomic forces.
Details on the hyperparameters that were optimized and the resulting set of optimal hyperparameters are given in Section S.5 of the Supplementary Information.. 

The hyperparameter optimization indicated the following ingredients to be important to obtain an accurate ACE-MLP.
First, we observed that the weight of the forces in the loss function, $\kappa$, needed to be reduced significantly, from typical values of 0.1 to 0.001.
We suspect that this helps to train weak dispersive interactions.
Second, we were able to increase the long-range cutoff to 7~\AA, likely reflecting the availability of more data in the training set for the final optimization.
Third, we found similar accuracy when H-H pair terms were used for the long-range interactions, but at an increased computational cost due to the higher number of hydrogen atoms.
We also found no improvement including  ``O-O'' and ``C-O'' pair terms in the long-range cutoff in addition to the ``C-C'' interactions. In short,  the ``C-C'' interaction was  sufficient to describe long-range interactions in the systems of interest.
In systems where polar and ionic interactions are more important, the framework could be extended to have multiple long-range cutoffs for different pair interactions, which we leave as a topic for future work.

The exploration of new configurations \textit{via} AL, addition of single molecules from MD and single molecule ``fragments'', and an additional hyperparameter optimization resulted in a stable MLP for all systems under investigation, exemplified by stable dynamics for configurations of 100~molecules for all systems investigated for 2~ns at different temperatures. 
In the C6ol energy profile, this manifests as a closer match to the DFT bond energy at high bond distances (see Fig. S4 of the Supplementary Information, dark purple curve).

%
The fact that a stable MLP for alkanes and alkanols and an adipate was obtained is evidence that adding explicit DFT energies and forces at large interatomic distances (\textit{i.e.}, ``fragments'') mitigated the problem of molecular fragmentation, and that the AL framework on its own could not produce a MLP that was robust. Adding simulations at more aggressive thermodynamics conditions (\textit{i.e.}, higher temperatures and pressures) could solve this issue and will be explored further in future work. 

\subsection{Property calculations}
\label{sec.methods.analysis}

We proceeded to the  calculation of properties for alkanes, alkanols, and the adipate. 
Since we were interested in the extrapolation of the ACE-MLP to longer-chain alkanes and alkohols, we additionally excluded C4 and C4ol from further analysis; instead, we analyzed the accuracy of the ACE-MLP for dodecane (C12) and dodecanol (C12ol), which the model had not seen during training.

\subsubsection{Vibrational frequencies}
\label{sec.methods.frequencies}
To assess the capacity and accuracy of the ACE-MLP to reproduce intramolecular interactions,
we calculated vibrational frequencies of single molecules in the gaseous state with the ACE-MLP and DFT, using the same parameters as described in~\autoref{sec.methods.DFT} for the latter method, except that
 4~\AA ~of separation were used between periodic images of molecules.
 We relaxed the atomic positions by minimizing the energy and forces independently for the ACE-MLP and DFT.
We used the conjugate gradient method for the relaxation with 0.001 eV\,\AA$^{-1}$ as the convergence criterion of the forces.
Next, we calculated the vibrational frequencies in the harmonic approximation using a finite-difference approach\cite{parlinski_first-principles_1997} implemented in \textit{MedeA}-\texttt{Phonon}.
The six lowest frequencies were discarded as the rotational and translational degrees of freedom.

\subsubsection{Density -- temperature relationship}
\label{sec.methods.density}
We generated systems containing 100 molecules of each alkane, alkanol, and adipate system with the \textit{MedeA} \texttt{Amorphous Materials Builder} at an initial density of 0.7 g\,cm$^{-3}$ and performed short relaxations with PCFF+.
Next, we performed NPT simulations with the ACE-MLP at 1~atm of pressure and at temperatures from 298~K to 318~K in steps of 5~K.
The MLP-ACE MD simulations were run for 2~ns, with a timestep of 0.5~fs, and configurations were saved every 1~ps.
Temperature and pressure were controlled with the Nos\'e--Hoover thermostat and barostat using temperature and pressure damping parameters of 100~fs.
The density was obtained from the mean density observed in the NPT dynamics of 2 ns, discarding the first 100~ps as thermalization and equilibration.
All ACE-MLP MD simulations were performed with LAMMPS.~\cite{LAMMPS}

\subsubsection{Isobaric heat capacity and vaporization enthalpy}
\label{sec.methods.heatcapacity}
The isobaric heat capacity of the liquid phase, $C_{P}$, is the sum of the isobaric heat capacity of the ideal gas, $C_{P}^{ideal}$, and the residual heat capacity, $C_{P}^{residual}$.
The frequencies in the molecule, calculated as described in \autoref{sec.methods.frequencies}, were used to evaluate the vibrational contribution to the isometric heat capacity.~\cite{mcquarrie_physical_1997}
Translational and rotational contributions were added according to McQuarrie and Simon~\cite{mcquarrie_physical_1997} to obtain the ideal isometric heat capacity, $C_{V}^{ideal}$.
$C_{P}^{ideal}$ was evaluated \textit{via} Mayer’s law,~\cite{cerdeirina_towards_2004} and
$C_{P}^{residual}$ was approximated for temperatures far below the critical temperature by: 
\begin{equation}
C_P^{residual}\sim(d\Delta H_{vap}/dT),
\label{eq.cp-res}
\end{equation}
where $\Delta H_{vap}$ is evaluated from:
\begin{equation}
\Delta H_{vap}= \langle E_{tot}^{gas} \rangle + RT - \langle E_{tot}^{liquid} \rangle .
\label{eq.Hvap}
\end{equation}
In the above equation, $\langle E_{tot}^{gas} \rangle $ is the mean total energy over 1~ns of NVT simulation of a single molecule in a (10~nm)$^3$ box,
$\langle E_{tot}^{liquid}\rangle$ is the mean total energy of the liquid equilibrated in the simulations, as described in \autoref{sec.methods.density}, divided by the number of molecules.
$R$ is the ideal gas constant, and $T$ is  the temperature set by the thermostat.

\subsubsection{Radial distribution function}
\label{sec.methods.rdfs}
We calculated the radial distribution function (RDF), 
$g(r)_{S-S'}$, of species $S$ with species $S'$ as:
\begin{equation}
 g_{S-S'}(r) = \frac{\rho(r)}{f(r)} = \frac{1}{f(r)} \frac{1}{N_S} \sum_i^S
\sum_j^{S'}\Big\langle \delta \left(r - \left| \bm r_i(t) - \bm r_j(t) \right| \right) \Big\rangle_t,
\label{eq:rdf}
\end{equation}
where $f(r)$ is the ideal-gas average number density at the same mean density, $\bm r_i(t)$ is the position of atom $i$ at time $t$, and
$\langle\cdots\rangle_t$ is a time average over the trajectory, equal to an ensemble average under the assumption of ergodicity.

\subsubsection{Diffusion coefficients}
\label{sec.methods.diffusion}

We calculated the tracer diffusion coefficient for carbon, $D_{tr}^\mathrm{C}$, from the mean-square displacement using the Einstein relation:~\cite{allen_computer_1987}
\begin{align}
\notag D_{tr}^\mathrm{C}=& 
\lim_{t \rightarrow \infty} \frac{1}{6t}  \left\langle  \mathrm{MSD}(t) \right\rangle_{NPT} \\
=&\lim_{t \rightarrow \infty}   \frac{1}{6t}\frac{1}{N_\mathrm{C}} \sum_{i=1}^\mathrm{N\mathrm{C}}  \left\langle |\bm r_i(t+\tau)-\bm r_i(\tau)|^2 \right\rangle_\tau,
\label{eq-Dtr}
\end{align}
where $\langle\cdots\rangle_{NPT}$ indicates an average in the NPT ensemble and 
 $\langle\cdots\rangle_\tau$ a time average in the trajectory, equivalent under the assumption of ergodicity. 
We fit the slope of the MSD between 200~ps and 300~ps, where the diffusion was linear for all systems while still returning good statistics, and employed block averaging over 4 independent blocks to calculate an error.


\subsection{Reference AIMD}
\label{sec.methods.aimd}

In order to benchmark the accuracy of our trained MLP, we performed \textit{ab initio} molecular dynamics (AIMD) simulations in the NPT ensemble at 1~bar and 318~K for both C6-6 and C6ol-6.
The same parameters were used for the electronic minimization as in \autoref{sec.methods.DFT}.
To reduce the computational cost, we employed a timestep of 1~fs and saved the positions every 10 steps.
We reached a simulation length of 90~ps for C6ol-6 and C6-6.
We discarded the first 10~ps for equilibration when calculating the RDF and density.
\FloatBarrier

\section{Results and discussion}
\label{sec:results}

\subsection{Vibrational frequencies and heat capacity of molecules}
\label{sec.res.freqs}
\begin{figure*}[t!]
    \centering
    \includegraphics[width=7.35in]{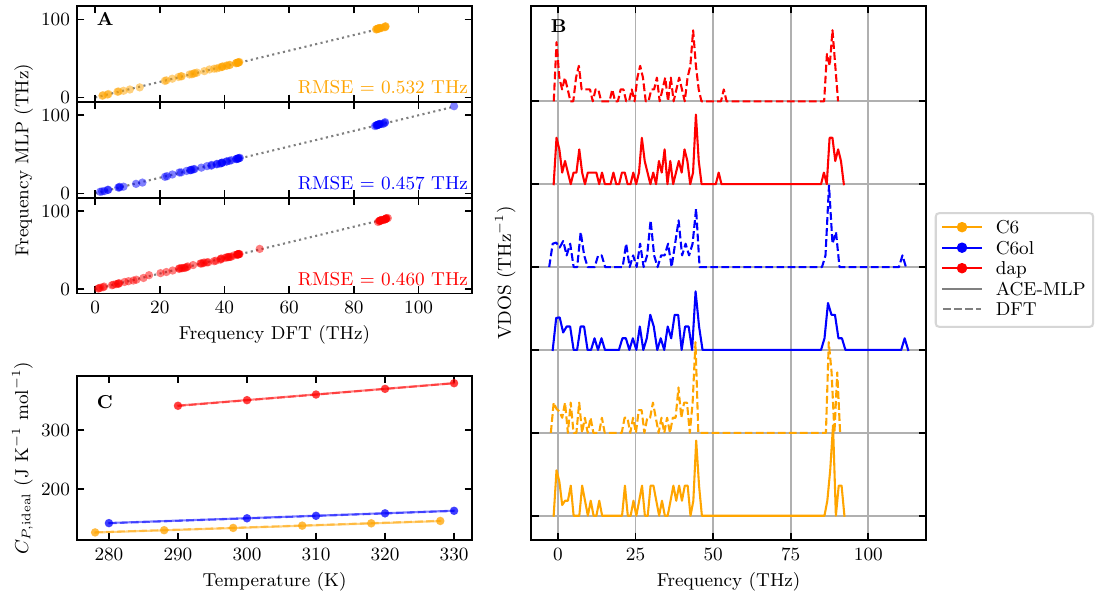}
    \caption{
    Analysis of the ACE-MLP accuracy for intramolecular interactions. C6 (orange),  C6ol (blue), and dap (red) are shown as representative for all alkanes and alkanols, resepctively, in the systems of interest. Properties are computed from the ACE-MLP (solid lines) and directly from DFT (dashed lines). 
    (A) The vibrational frequencies. The RMSE is reported in the lower left for each system. The dotted gray line has a slope of 1 as a guide. (B) The vibrational density of states (VDOS). Spectra are vertically offset for clarity. 
    (C) The ideal isobaric heat capacity, $C_p^{ideal}$, computed from the vibrational frequencies.
    }
    \label{fig:res-frequencies}
\end{figure*}

To assess the accuracy of intramolecular interactions in the trained ACE-MLP, we calculated 
the vibrational frequencies (see \autoref{sec.methods.frequencies}), the vibrational density of states (VDOS), and $C_p^{ideal}$ for C6, C6ol, and dap as representatives of each class of materials under investigation.
The ACE-MLP computed frequencies for these molecules are shown in \autoref{fig:res-frequencies}A
against the frequencies computed directly with DFT.
We observed a root mean square error (RMSE) below 1~THz between the ACE-MLP frequencies and the DFT frequencies, indicating excellent agreement. 
The remarkable reproduction of frequencies also resulted in an excellent agreement in the vibrational density of states (VDOS), as shown in \autoref{fig:res-frequencies}B.
In turn, this allowed for good agreement for the $C_p^{ideal}$ between the ACE-MLP and DFT, as shown in \autoref{fig:res-frequencies}C.
We concluded that the PES of the single molecules and, therefore, intramolecular interactions were well-described by the ACE-MLP.

\subsection{Densities}

\begin{figure}[th]
    \centering
    \includegraphics[width=3.54in]{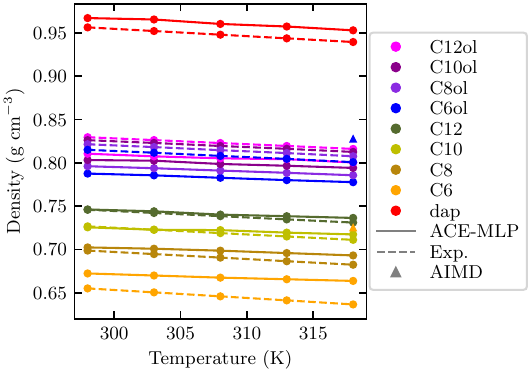}
    \caption{
    A comparison of the densities as a function of temperature for all systems calculated with the ACE-MLP (solid lines) against the experimental values (dashed lines). 
    Experimental densities for alkanes were taken from NIST ~\cite{informatics_nist_nodate}, for alkanols were taken from Matsuo \textit{et al.}~\cite{matsuo_volumetric_1989} and for the adipate from Diogo \textit{et al.}\cite{diogo_density_2014}
    For C6 and C6ol at 318~K, the densities obtained from AIMD simulations (triangles) are also shown.
    }
    \label{fig:res-densities}
\end{figure}

In order to assess the quality of intermolecular interactions with the ACE-MLP, we evaluated the densities obtained in NPT simulations, varying the temperature at a constant pressure of 1~atm (see \autoref{sec.methods.density}).
The densities obtained with the ACE-MLP at each temperature are shown in \autoref{fig:res-densities} for all molecules, along with the experimental densities.~\cite{informatics_nist_nodate, matsuo_volumetric_1989}

For alkanes, the density was accurately reproduced for the longer-chain alkanes (\textit{e.g.}, C12 had an average relative error of $0.7\pm0.2$\%), but the accuracy was slightly worse for shorter-chain alkanes (\textit{e.g.}, C6 had an average relative error of $4.3\pm0.7$\%). The ACE-MLP tended to overestimate the densities at elevated temperatures for an average relative error of $1.3\pm1.3$\% over all alkanes.
For the alcohols, the densities were slightly and systematically underestimated with an average relative error of $2.6\pm0.4$\%.
Nevertheless, the relative increase in density with chain length was reproduced by the ACE-MLP for both molecule types.
For the dap system, the error in the density was 1.3\%.
The maximum relative error in the density at any temperature was 4.3\% for alkanes (C6 at 318~K) and 3.3\% for alcohols (C6ol at 298~K), which is consistent with results by Magd\u{a}u \textit{et al.}~\cite{magdau_machine_2023} 
We concluded that long-range interactions are accurately captured with the split descriptor, whether the underlying architecture is GAP, as in Magd\u{a}u \textit{et al.},~\cite{magdau_machine_2023} or ACE, demonstrated in this work. 

The remaining discrepancy in the densities for the ACE-MLP  could be due to either the accuracy of the DFT or the accuracy of the ACE-MLP with respect to DFT.
To differentiate between these two cases, we ran AIMD simulations of systems of C6 and C6ol (see \autoref{sec.methods.aimd}) at 318~K.
The densities computed from these simulations (\autoref{fig:res-densities}, triangles) 
were not conclusive, as the densities for both C6-6 and C6ol-6 were significantly overestimated (relative errors of 13.7\% and 4.6\%, respectively), most likely due to small cell sizes and insufficient simulation time in the AIMD runs.
We note that the densities computed using the ACE-MLP were more accurate than the densities from AIMD simulations, even though the MLP was trained on DFT data.
Because the density, as with many other properties, was estimated from larger ensembles from MD simulations, we reasoned the better statistical representation from the ACE-MLP simulations offset possible inaccuracies in the representation of the PES.

\subsection{Radial distribution function}

\begin{figure}[t!]
    \centering
    \includegraphics[width=3.54in]{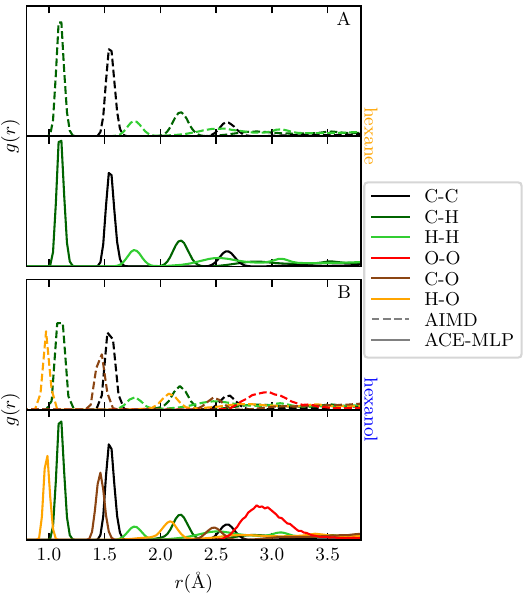}
    \caption{A comparison of the RDFs for C6 (A) and C6ol (B) obtained from the ACE-MLP (solid lines) and from AIMD simulations (dashed lines) at 318~K. The RDFs from the AIMD simulations are vertically offset for clarity.
    }
    \label{fig:RDFs}
\end{figure}

To further evaluate the accuracy of the ACE-MLP, we computed the RDFs (see \autoref{sec.methods.rdfs}) for each pair of atom-types from the ACE-MLP and the AIMD simulations. 
The RDF for C6 and C6ol are displayed in \autoref{fig:RDFs}, showing good agreement between the ACE-MLP and AIMD.
All peaks in the RDFs for C6 and C6ol matched those from AIMD, with only a slight difference in height for the ``O-O'' peak in C6ol.
We suspect that discrepancy was due to the smaller number of molecules in the AIMD simulations, which constrained the O-H bonding network and yielded a lower number of observations in the simulation.
Despite this discrepancy, the ``O-O'' and ``O-H'' RDFs demonstrate that the ACE-MLP was able to properly account for hydrogen bonding interactions in the system.

\subsection{Diffusion coefficients}
\begin{figure}
    \centering
    \includegraphics[width=3.54in]{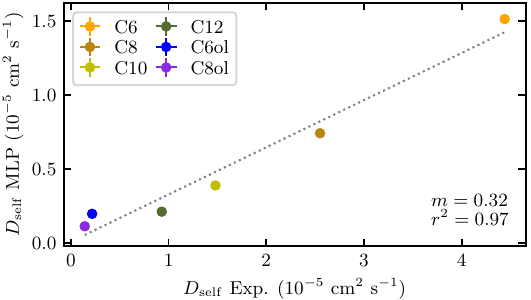}
    \caption{A comparison of the self-diffusion coefficients computed from the ACE-MLP (y-axis) against the experimentally known values. The experimental values were taken from Iwahashi \textit{et al.}~\cite{iwahashi_dynamical_1986, iwahashi_dynamical_1990}
    The least-squares fit is shown as a dashed grey line, and the slope and $r^2$ of the fit are reported in the bottom left corner. 
    C10ol, C12ol, and dap are omitted due to lack of experimental data.}
    \label{fig:res-diffusion}
\end{figure}

As a final test of the accuracy of the intermolecular interactions of the ACE-MLP, we computed the self-diffusion coefficients (see \autoref{sec.methods.diffusion}) and compared them against the experimentally known values~\cite{iwahashi_dynamical_1986, iwahashi_dynamical_1990} in \autoref{fig:res-diffusion}. Experimental values for C10ol, C12ol, and dap were not available in the literature, and those systems were omitted from this analysis.
We observed qualitative agreement where the diffusion coefficients decreased as chain length increased; however, the computed diffusion coefficients were consistently underestimated by a factor of $\sim \frac{1}{3}$.

Simulated self-diffusion coefficients often exhibit larger deviations from experimental values compared to static properties like density. 
For instance, Kulschewski and Pleiss\cite{Kulschewski_2013} noted that the self-diffusion coefficients calculated for 13 aliphatic alcohols using OPLS~2013 deviated from their experimental counterparts by as much as 55\%.
More recently, Allers \textit{et al.}\cite{Allers_2021} computed the self-diffusion coefficients for a set of 102 pure liquids using interaction parameters from the Generalized Amber Force Field (GAFF) along with RESP/B3LYP/6-31G* partial atomic charges.
They found that for some compounds, the deviation could exceed 50\%.
Methodological aspects, such as the size, duration and the choice of the statistical ensemble, can account for a small percentage of these deviations, as reviewed by Maggin \textit{et al}.\cite{Maginn_Messerly_Carlson_Roe_Elliot_2018}
However, to address larger deviations, the parametrization of the classical force field is typically modified, for example by adjusting partial atomic charges,\cite{Kulschewski_2013}
which is challenging for MLPs.
We suspect that the systematic underestimation of diffusion coefficients could be either due to the MLP not capturing accurately the barriers to diffusion of the DFT functional, or that the DFT functional underestimates barriers. Since our  choice of DFT functional was based on evaluations of static properties, a dynamic property such as the friction between two molecules might be poorly represented. 

Nevertheless, the results are encouraging for several reasons. First, calculating diffusion, an activated process, requires capturing the barrier to diffusion with high accuracy.\cite{van_der_ven_first-principles_2001}
Even a correct order-of-magnitude estimation of diffusion can be an improvement for many applications.
Second, the correct relative changes in diffusion with chain length allow the model to be reliably used in applications where the ranking of materials by diffusion is more important than the absolute value, as for high-throughput screening applications.\cite{kahle_high-throughput_2020}

\subsection{$C_p$ and $\Delta H_{vap}$}
\label{sec.res.cp}

\begin{figure*}[t!]
    \centering
    \includegraphics[width=7.35in]{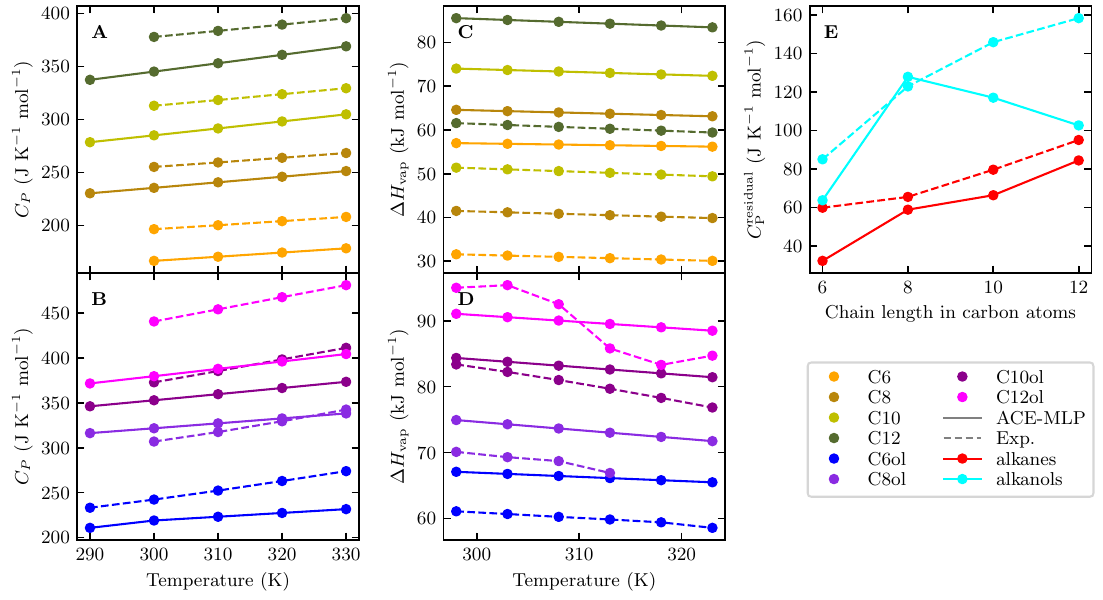}
    \caption{Analysis of the accuracy of the ACE-MLP (solid lines) for thermodynamic property prediction for all molecules in the systems of interest except for dap, for which no experimental data for thermodynamic properties was found.
    (A-B) The isobaric heat capacities, $C_P$. 
    (C-D) The enthalpies of vaporization, $\Delta H_{\rm{vap}}$. 
    (A, C) Experimental results (dashed lines) were taken from NIST.~\cite{informatics_nist_nodate} 
    (B) Experimental results (dashed lines) were taken from Z\'abransk\'yr \textit{et al.}~\cite{zabransky_heat_1990} 
    (D) Experimental results (dashed lines) were taken from Kulikov \textit{et al.},~\cite{kulikov_enthalpies_2001} Nasirzadeh \textit{et al.},~\cite{nasirzadeh_vapor_2006} Pokorn\'y \textit{et al.},~\cite{pokorny_vapor_2021} and NIST~\cite{informatics_nist_nodate} for C6ol, C8ol, C10ol, and C12ol, respectively.
    (E) The $C_p^{residual}$ as a function of the chain length for alkanes (orange) and alcohols (blue).
    The experimental results (dashed lines) were calculated from the the data in C,D.
    }
    \label{fig:thermo}
\end{figure*}

Next, we evaluated the accuracy of the ACE-MLP on thermodynamic properties by calculating the $C_{P}$ (see \autoref{sec.methods.heatcapacity}) of each species in the systems of interest.
These results are presented against the experimental values~\cite{informatics_nist_nodate, zabransky_heat_1990, kulikov_enthalpies_2001, nasirzadeh_vapor_2006, pokorny_vapor_2021} in \autoref{fig:thermo}A,B. 
Because experimental properties for dap were found in the literature, we excluded that system from this analysis.
Similar to the diffusion coefficients above, we observed that the ACE-MLP was able to qualitatively predict the $C_{P}$ for each system. 
The average deviation to experimental data over all molecules was $9.4\pm4.1$\%, with a maximum deviation (C12ol) of 16.0\%.
This agreement is encouraging and
most likely due to the strong correlation observed between the ACE-MLP and DFT in the vibrational frequency calculations (see \autoref{sec.res.freqs}). 
For reference, in C6 and C6ol, the vibrational contribution represents $\sim 85$\% of the total $C_{P}^{ideal}$.

The intermolecular contribution to $C_{P}$ depends on an accurate $d\Delta H_{vap}/dT$ relationship (\autoref{eq.cp-res}), which we plot in \autoref{fig:thermo}C,D. 
For the alkanes (\autoref{fig:thermo}C), we observed that, 
over the temperature range studied in this work, the calculated $\Delta H_{vap}$ overestimated the experimental data by 45-83\%.
For the alkanols (\autoref{fig:thermo}D), we observed a slight overestimation.
Again, similar to the density calculations, the overestimation was larger for the shorter-chain molecules for both alkanes and alkanols. Taken together, these observations could imply that the interactions due to the terminal $\text{CH}_3$ groups were overestimated to a greater extent than the interactions stemming from $\text{CH}_2$ groups.
At temperatures further from the critical temperature, the thermal dependence of $\Delta H_{vap}$ was considered linear.
In all cases, a linear fit resulted in an $R^2 > 0.97$.
We plot the evolution of calculated and experimental $C_{P}^{residual}$ with the number of carbon atoms of the alkyl chains in \autoref{fig:thermo}E.
For alkanes (red), while the $C_{P}^{residual}$ was overestimated for C6, the experimental trend was reproduced by the ACE-MLP.
However for the alkanols (cyan), the values calculated for C10ol and C12ol diverged significantly from the experimental data.
This behavior may be due to an underrepresentation of the OH-OH interaction in the training set, and therefore reduced accuracy in the ACE-MLP. 
To confirm this suspicion, we evaluated the energies of two alcohols as a function of OH-OH separation, and we 
observed that the minimum of the potential was at shorter distance in the ACE-MLP than in the DFT calculations.
This observation is discussed in greated detail in 
Section S6 of the Supplementary Information 
and shown in Fig. S5. 

Predicting isobaric heat capacities from atomistic MD simulations is notoriously difficult, as confirmed in the literature.
Zhang \textit{et al.}~\cite{zhang_tiny_2022} noticed that the isobaric heat capacities of liquid alkanols calculated directly from the slope of a linear fit to enthalpies versus temperature are twice as large as the experiments using the OPLS-AA force field.
The same conclusion was drawn by Zahariev and Ivanova~\cite{zahariev_molecular_2013} who calculated the isobaric heat capacity of liquid alkanes from enthalpy fluctuations using the AMBER99 force field.
It is a well known problem that all-atom force fields overestimate the vibrational energy due to nuclear quantum effects.\cite{gao_comparing_2021}
However, nuclear quantum effects are less significant for the calculation of  $\Delta H_{vap}$.
Indeed, following a similar approach as employed in this work, Zahariev \textit{et al.}~\cite{zahariev_fully_2014} obtained a maximum deviation for $C_p$ of $\sim 15$\% relative to experiments with the AMBER99 force field for alkanes. 
We note that the level of accuracy obtained with the ACE-MLP is similar to that obtained with a classical force field carefully refined on experimental data.
To further improve $C_p$ predictions, improvements in both interatomic potentials and the evaluation of ideal-gas vibrational heat capacities, as highlighted by Stejfa \textit{et al.},~\cite{stejfa_first-principles_2019} should be explored.

\subsection{Evaluation of the dual-descriptor MLP}
In order to showcase the validity of the dual descriptor, we used the same hyperparameters as for the ``best'' ACE-MLP and trained several single-descriptor MLPs at different cutoffs. Each single-descriptor MLP contained the same number of model parameters.
We evaluated the resulting MLPs on two criteria: 1) a stability score, $R_s$, defined by the number of valid configurations reached for all alkane and alkanol systems (298~K to 318~K) when running with NPT for 2~ns (sampling every 1~ps) divided by the target number of configurations, and 2) the error in the density with respect to experiment.
As the cutoff in the single-cutoff models was increased from 3~\AA ~to 7~\AA, the stability score $R_s$ dropped steadily from 1.00 ($r_c=3$~\AA) to 0.985  ($r_c=4$~\AA), 0.381 ($r_c=5$~\AA), 0.145 ($r_c=6$~\AA), and 0.014 ($r_c=7$~\AA) demonstrating that increasing the cutoff incurs a cost in accuracy and model robustness.
The loss of accuracy and robustness with increasing cutoff is likely due the increasing complexity of the PES to be learned by the MLP as the number of atoms within the cutoff radius increases.
For the only stable single-cutoff MLP ($r_c=3~\AA$), the calculated density had an average error of $4.3\pm2.5$\%, more than double the error obtained with the ``dual'' ACE-MLP ($2.0\pm1.2$\%). 
We highlight that the energies and forces, always evaluated on the same test set, were in good agreement with the DFT for all MLPs (see Figures~S6--S9 in the Supplementary Information).
These observations indicate that for single-cutoff MLPs there exists a delicate compromise between model stability and the quality of intermolecular interactions.
While a single-cutoff MLP  needs to navigate this compromise carefully, a dual-cutoff descriptor circumvents this compromise, allowing for both stable dynamics and accurate predictions of the bond interactions.

\section{Summary and Conclusions}
\label{sec:conclusions}

We have presented an approach to using uncertainty-guided AL to train an ACE-MLP for the prediction of the PES of organic liquids.
In order to run stable simulations of hundreds of molecules, the uncertainty-guided AL was enriched with the detection of bond-breaking events, and the training was augmented with high-energy single molecules with highly stretched bonds (\textit{i.e.}, fragments). 
These two ingredients proved essential in achieving a MLP capable of stable MD simulations,  demonstrating that particular care must be taken during training set generation to ensure the MLP remains in an interpolative regime during inference. Whether this AL framework can be fully automated (\textit{i.e.}, without human intervention) to the same level of stability will be investigated in future work.

The trained MLP inherently, without any additional terms, predicted both intramolecular and intermolecular energies and forces with high accuracy. 
As a result, it provided a robust tool for property predictions.
Importantly, no empirical calibration of model parameters was performed, demonstrating the utility of MLPs for describing systems for which there are not readily available empirical force field parameters or experimental data.

We find particularly noteworthy that in addition to the high accuracy obtained in energies and forces, stable dynamics could be performed, despite the MLP having been trained on a comparatively small training set of 2641 structures.
While there is no direct comparison for the same systems available in the literature, 935 structures were used by Magd\u{a}u \textit{et al.}~\cite{magdau_machine_2023} for mixtures of ethylene carbonate and ethyl-methyl carbonates, and 35,000 structures were used by Abedi \textit{et al}~\cite{abedi_high-dimensional_2023} for methane. This underscores the wide range of numbers of training configurations required to train an accurate MLP, depending on the complexity of the task to learn.
Our results highlight that complex systems can be described in a data-efficient manner, without the need for tedious calibration.

We demonstrated that properties like density could be predicted with an error below 4.3\% for
a wide range of aliphatic alcohols and linear alkanes.
The remaining errors are likely due to approximations of the underlying DFT method, as the ACE-MLP computed RDFs showed excellent agreement with AIMD.

Thus, the present work demonstrates that:
\begin{enumerate}
    \item Machine-learned potentials with a dual cutoff are capable of describing organic liquids and reflect the predictive power, generality, and accuracy of the underlying \textit{ab initio} method used in its training.  While message-passing MLPs with single cutoffs reach similar accuracy, employing a MLP without message-passing has advantages due to simpler training and better scalability in application.
    \item  We provide additional evidence for the well-documented need to improve the accuracy of DFT methods for describing intermolecular interactions, in order to further increase the accuracy with respect to experiment.\cite{eyert_machine-learned_2023, burke_perspective_2012, liu_phase_2022, tsuzuki_accuracy_2020}

\end{enumerate}
In conclusion, the stage is set for employing MLPs on large-scale MD simulations of systems containing organic liquids.
Further applications could include organic liquids in contact and reacting with solids, \textit{e.g.}, in areas such as heat transfer, tribology, catalysis, corrosion, energy conversion, and energy storage.

\section*{Acknowledgements}
The authors thank the ACE team, in particular Yury Lysogorskiy for their help with the ACE descriptor and the \texttt{pacemaker} code.
We also thank Volker Eyert, David Rigby, David Reith, Clive Freeman, and Brian Dron for helpful discussions and feedback on the manuscript.
We acknowledge the EuroHPC Joint Undertaking for awarding this project access to the EuroHPC supercomputer LUMI, hosted by CSC (Finland) and the LUMI consortium through a EuroHPC Regular Access call.

\bibliographystyle{rsc} 
\bibliography{bibliography} 
\newpage

\onecolumn
\newpage
\setcounter{section}{0}
\setcounter{figure}{0}
\renewcommand{\thesection}{S\Roman{section}}
\renewcommand{\thefigure}{S\arabic{figure}}
\renewcommand{\thetable}{S\arabic{table}}

\begin{center}
    \LARGE{\textbf{A dual-cutoff machine-learned potential for condensed organic systems obtained \textit{via} uncertainty-guided active learning -- Supplementary Information}}\\
    \vspace{0.3cm}
    \noindent\large{Leonid Kahle,$^{\ast}$\textit{$^{a}$} Benoit Minisini,\textit{$^{a
    }$}
    Tai Bui,\textit{$^{b}$}
    Jeremy T. First,\textit{$^{c}$}
    Corneliu Buda,\textit{$^{b}$}
    Thomas Goldman,\textit{$^{b}$}
    and Erich Wimmer\textit{$^{a}$}}
    \end{center}

    \footnotetext{\textit{$^{a}$~Materials Design SARL, 42 avenue Verdier, 92120 Montrouge, France}}
    \footnotetext{\textit{$^{b}$~bp Exploration Operating Co. Ltd, Chertsey Road, 
            Sunbury-on-Thames TW16 7LN, UK}}
    \footnotetext{\textit{$^{c}$~bp, Center for High Performance Computing, 225 Westlake Park Blvd, Houston, TX 77079, USA}}


\subsection{Choice of DFT functional}
\label{sec:supp.sec.resultsdft}

\begin{figure*}[h!]
    \centering
    \includegraphics[width=0.8\hsize]{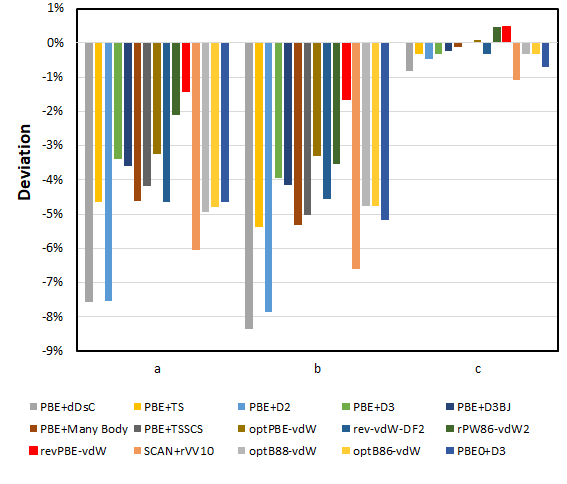} \\
    \caption{Deviation between DFT (0 K) and experimental (4 K)~\cite{avitabile_low_1975} cell parameters (a,b, c) of crystalline polyethylene for different exchange correlation functionals.}
    \label{fig:PE}
\end{figure*}

\begin{figure*}[h!]
    \centering
    \includegraphics[width=0.8\hsize]{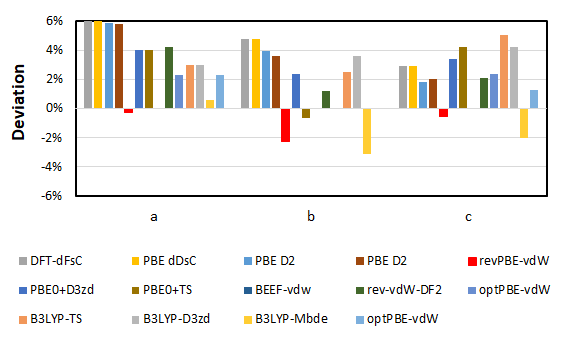} 
    \caption{Deviation between DFT (0 K) and experimental (122 K)~\cite{kirchner_cocrystallization_2008} cell parameters (a,b, c) of crystalline methanol for different exchange correlation functional.}
    \label{fig:methanol}
\end{figure*}

\begin{figure}[h!]
    \centering 
    \includegraphics{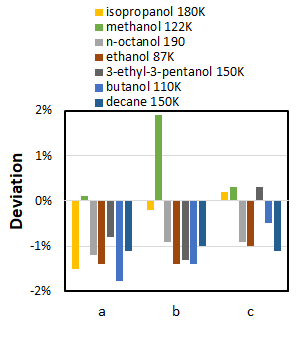} \\
    \caption{Deviation between DFT (0 K) and experimental cell parameters (a,b, c) of crystalline organic materials, isopropanol,~\cite{ridout_low-temperature_2014} methanol,~\cite{kirchner_cocrystallization_2008} n-octanol,~\cite{shallard-brown_n-octanol_2005} ethanol,~\cite{jonsson_hydrogen_1976}  3-ethyl-3-pentanol,~\cite{bond_3-ethylpentan-3-ol_2006} butanol,~\cite{derollez_structure_2013} and decane.~\cite{bond_n-decane_2002} }
    \label{fig:all_crystal}
\end{figure}

Our calculations on polyethylene and methanol showed revPBE-vdw yielded the lowest deviation from experimental data (\autoref{fig:PE} and \ref{fig:methanol}), with a maximum absolute deviation of 2.2\% for the $b$ lattice constant of methanol.
The mean absolute deviation between the revPBE-vdw functional and experimental results was  $1\pm 0.8$\% for lattice constants calculated for the six other organic molecular crystals (\autoref{fig:all_crystal}).
VASP was used for all calculations with settings as for the training calculations and a k-point spacing of 0.5~\AA$^{-1}$.

\newpage
\setlength{\extrarowheight}{10pt}
\begin{table*}
    \centering
    \begin{tabular}{cp{7cm}p{4cm}ccc}
   
    \toprule
     \textbf{Step} & \textbf{Systems added} &  \textbf{Exploration Technique}   & Conf. added & Conf. total \\[8pt]
    \midrule
    0 & \underline{250~C4ol-8}, \underline{50~C4ol-1} & Expl. pcff+ and rand. displ. &  300 & 300\\[8pt]
      1 & \textbf{70~C4ol-8}, \underline{100~C6ol-6}, \underline{100~C8ol-4}, \underline{100~C10ol-3} & Expl. MLP and pcff+  & 370 & 670 \\[8pt]
    2 & \textbf{19~C4ol-8}, \textbf{23~C6ol-6}, \textbf{13~C8ol-4}, \mbox{\textbf{40~C10ol-3}} & Expl. MLP & 95 & 765 \\
    3 & \textbf{37~C4ol-8, 31~C6ol-6, 26~C8ol-4, 26~C10ol-3} & Expl. MLP &  120 & 885 \\
    4 & \textbf{31~C4ol-8, 18~C6ol-6, 13~C8ol-4, 10~C10ol-3} & Expl. MLP & 72 & 957\\
    5 & \textbf{60~C4ol-8, 18~C6ol-6, 32~C8ol-4, 73~C10ol-3} & Expl. MLP & 183 & 1140 \\
    6 & \textbf{23~C4ol-8, 7~C6ol-6, 9~C8ol-4, 27~C10ol-3} & Expl. MLP &66 & 1206 \\
    7 & \textbf{12~C4ol-8, 5~C6ol-6, 5~C8ol-4, 4~C10ol-3, 2~C6-6, 1~C10-3} & Expl. MLP &29 & 1235\\
    8 & \textbf{6~C4ol-8, 7~C6ol-6, 3~C8ol-4, 3~C10ol-3, 1~C10-4}, \underline{90~dap-1}, \underline{440~dap-3} & Expl. MLP and pcff+ &550 & 1785 \\
    9 & \textbf{1~C4ol-8, 1~C6ol-6, 1~C10-4, 26~dap-3} & Expl. MLP &29 & 1814 \\
    10 & \textbf{4~C4ol-8, 4~dap-3} & Expl. MLP &8 & 1822 \\
    11 & \textit{18~C4ol-1}, 
         \textbf{5~C4ol-8,}
         \textit{50~C6ol-1},
         \textit{50~C8ol-1},
         \textit{50~C10ol-1},
         \textit{50~C6-1},
         \textit{50~C10-1},
         \textit{50~dap-1},
         \textbf{22~dap-3} & Expl. MLP + fragments &345 & 2167\\
    12 & 16~C4ol-1,
         \textbf{4~C4ol-8,}
         \textit{50~C6ol-1},
         \textit{50~C8ol-1},
         \textbf{1~C8ol-4},
         \textbf{1~C10ol-3,}
         \textit{50~C10ol-1},
         \textit{50~C6-1},
         \textit{50~C10-1},
         \textbf{47~C10-4,}
         \textit{50~dap-1},
         \textbf{13~dap-3} & Expl. MLP + fragments &382 & 2549\\
    13 & \textbf{10~C4ol-8, 13~C6ol-6, 16~C8ol-4, 10~C10ol-3, 16~dap-3} & Expl. MLP &65 & 2614\\
    14 & \textbf{7~C4ol-8, 1~C8ol-4, 3~C10ol-3, 7~dap-3, 50~C4-2ol-6} & Expl. MLP + pcff+ &68 & 2682 \\
    15 & \textbf{3~C4ol-8, 12~C6ol-6, 11~C8ol-4, 9~C10ol-3, 1 C4-8, 5~C6-6, 9~C10-4, 16~dap-3} & Expl. MLP &66 & 2748 \\
    16 & \textbf{3~C4ol-8, 7~dap-3} & Expl. MLP &10 & 2758 \\
    17 & \textit{127~dap-1, 24~C6-1, 33~C6ol-1} & fragments & 183 & 2941 \\
    18 & & cleaning & -300 & 2641 \\
    \bottomrule
    \end{tabular}
    \caption{Structures added during the active learning and training set configurations at every iteration (step). Configurations created using PCFF+ driven molecular dynamics or random displacements are underlined, configurations originating from the ``fragmentation'' algorithm (stretching single molecules) are in italics, and configurations from the uncertainty-guided exploration \textit{via} the MLP are given in bold.}
    \label{tab.sup.alres}
\end{table*}

\twocolumn

\subsection{Active Learning bond potential progression}
\label{sec.sup.training}
\begin{figure}[b!]
    \centering
    \includegraphics[width=3.54in]{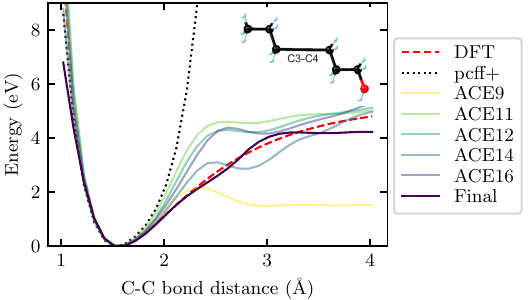}
    \caption{The energy profile of C3-C4 carbon bond vector (see insert for illustration) of C6ol at various stages of training for the ACE-MLP.
    PCFF+ is shown as a dotted black curve, the DFT results are shown as a dashed red line, and the MLP results at various iterations of the AL cycle are shown as solid colored lines. The final curve represents the ACE-MLP after all AL cycles and hyperparameter optimization.
    }
    \label{fig:nrj_fragm}
\end{figure}

In \autoref{fig:nrj_fragm} we show the energy of an hexanol molecule as a function of the C3--C4 separation while all other bond distances are kept constant.
PCFF+ is shown as a black curve with points, as a representative of a classical bonded force field, displaying the quartic behavior of the bonding term.
As expected, the DFT results (dashed red line) displayed a steep increase in energy as the bond distance was reduced below the equilibrium distance and a shallow, but steady, increase as the molecule was pulled apart towards a fragmentation energy of roughly 5~eV.

The MLP at different stages of the AL cycle is shown in the solid colored lines in \autoref{fig:nrj_fragm}. At cyle 9, we observed that the MLP had the wrong fragmentation energy and a small local maximum at 2.5~\AA , most likely due to the lack of training data in this region. 
After the introduction of fragments in cycle 11, the correct fragmentation energy was recovered, and the local maximum dissipated. The same local maximum was still present to a lesser extent in cycle 16, but the fragmentation energy was better reproduced, which we attribute to the inclusion of fragments in the training configurations. After refining the training set configurations, the final MLP displayed the expected behavior.
The final curve in solid dark purple represents the ACE-MLP after all AL cycles, training set refinement, and hyperparameter optimizations. 
This curve closely results in good qualitative agreement with the DFT energy profile and no energy barrier, only deviating from the DFT energy curve at unphysically large bond lengths (\textit{i.e.}, >3 \AA), indicating an accurate prediction of the bonding interaction.

\subsection{Definition of robustness in MLP exploration}
\label{sec.sup.robustness}
The minimal dump (or sampling) period is one configuration every 40~fs, the maximum is one configuration every 500~fs.

A success ratio $R_S$ is defined as the number of stable and non-extrapolated structures, divided by the number of target structures.
A structure is stable if:
\begin{itemize}
    \item No unphysical bond distances are present in the configuration, which are bond distances below 0.6~\AA ~or above 2.6~\AA.
    \item $\gamma_{max}$ is below 100, as very high extrapolation grades are a sign of unstable configurations.
\end{itemize}
A structure is extrapolated if $\gamma_{max} > 1$.

Our goal was to keep this ratio at $\approx 0.5$, in order to balance exploring new configurations without pushing the MLP into an extremely extrapolative regime.
Therefore, we set a lower threshold at 0.25 and a higher threshold at 0.75. If the $R_S$ dropped below 0.25, the algorithm reduced the sampling period by a factor of 2. If the $R_s$ rose above 0.75, the algorithm doubled the sampling period.

\subsection{Active Learning}
\label{sec.supp-al}

A typical input file to \texttt{pacemaker} used during the AL is:
\begin{verbatim}
cutoff: 6.0
seed: 1
metadata:
  origin: Automatically generated input
potential:
  deltaSplineBins: 0.001
  elements:
  - C
  - O
  - H
  embeddings:
    ALL:
      npot: FinnisSinclairShiftedScaled
      fs_parameters:
      - 1
      - 1
      - 1
      - 0.5
      ndensity: 2
  bonds:
    ALL:
      radbase: SBessel
      radparameters:
      - 5.25
      rcut: 3.0
      dcut: 0.1
      NameOfCutoffFunction: cos
      r_in: 0.4
      delta_in: 0.4
      core-repulsion:
      - 5.0
      - 5.0
    CC:
      radbase: SBessel
      radparameters:
      - 5.25
      rcut: 6.0
      dcut: 0.1
      NameOfCutoffFunction: cos
      r_in: 0.4
      delta_in: 0.4
      core-repulsion:
      - 5.0
      - 5.0
  functions:
    ALL:
      nradmax_by_orders:
      - 15
      - 6
      - 4
      - 2
      lmax_by_orders:
      - 0
      - 3
      - 2
      - 1
    number_of_functions_per_element: 500
data:
  filename: step-12-label.pckl.gzip
  test_size: 0.1
fit:
  loss:
    kappa: 0.001
    L1_coeffs: 0
    L2_coeffs: 1.0e-08
  optimizer: BFGS
  maxiter: 500
backend:
  evaluator: tensorpot
  batch_size: 140
  display_step: 100                         
\end{verbatim}

\subsection{Hyperparameter optimization}
\label{sec.sup.hyperopt}

We performed for a hyperparameter optimization \textit{via} an extensive grid search. The following hyperparameter values were explored:
\begin{itemize}
    \item $\kappa$: 0.1, 0.01, 0.001
    \item $l_2$: 1e-6, 1e-7
    \item $n_m$ ($l_m$): 25,5,3 (0,3,2); 25,15,5  (0,3,2); 16,8,4,2 (0,3,2,1)
    \item $r_c^s$: 3, 4
    \item $r_c^l$: 6, 7, 8
    \item number of functions per element: 600, 800, 1000, 1250, 1500
    \item Long distance bond: ``H-H'', ``C-C''
    \item nradmax / lradmax of 16,8,4,2/0,3,2,1 \& 25,5,3/0,3,2
\end{itemize}

The optimal set of hyperparameters had a short range cutoff $r_c^s$ at 3~\AA, a long range cutoff $r_c^l$ at 7~\AA, a force weight $\kappa$ of 0.001, 1500 independent functions per element, and the long distance interaction was on the C-C bond. 
We also saw that it is worth constraining the expansion of ACE to 4-body terms, and providing more functions to 2 body terms than the default settings.
The input file employing the optimal set of hyperparameters is:
\begin{verbatim}
cutoff: 7.0
seed: 1
potential:
  deltaSplineBins: 0.001
  elements:
  - C
  - H
  - O
  embeddings:
    ALL:
      npot: FinnisSinclairShiftedScaled
      fs_parameters:
      - 1
      - 1
      - 1
      - 0.5
      ndensity: 2
  bonds:
    ALL:
      radbase: SBessel
      radparameters:
      - 5.25
      rcut: 3.0
      dcut: 0.01
    CC:
      radbase: SBessel
      radparameters:
      - 5.25
      rcut: 7.0
      dcut: 0.01
  functions:
    ALL:
      nradmax_by_orders:
      - 25
      - 15
      - 5
      lmax_by_orders:
      - 0
      - 3
      - 2
    number_of_functions_per_element: 1500
data:
  filename: labels_step_all-maxfor-10.pckl.gzip
  test_size: 0.02
fit:
  loss:
    kappa: 0.001
    L1_coeffs: 0
    L2_coeffs: 1.0e-06
  optimizer: BFGS
  maxiter: 1500
backend:
  evaluator: tensorpot
  batch_size: 512
  display_step: 100
\end{verbatim}


\begin{figure*}[t!]
    \centering
    \includegraphics[width=7.35in]{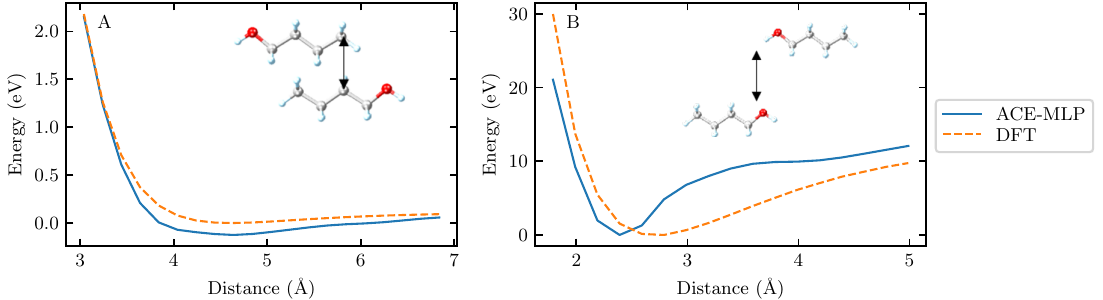}
    \caption{Displacement energy profiles for two different configurations of interacting C4ol molecules. 
    Insets illustrate the distance vector with a black line. 
    The ACE-MLP results are shown as the solid blue lines, and the DFT results are shown as the dashed orange lines.
    (A) The energy of system was computed along the chain to chain (carbon to carbon) interaction.
    (B) The energy of system was computed along the OH-OH (hydrogen to oxygen) interaction. 
    }
    \label{fig:nrj_configs}
\end{figure*}

\subsection{Investigation of the OH-OH interaction}
\label{sec.sup.configurations}

To test how accurately intermolecular interactions were reproduced by the MLP, we created two configurations of butanol dimers, as displayed in \autoref{fig:nrj_configs} (insets).
The chain-chain configuration contained two molecules placed in parallel along the chain, with the OH-groups as far apart as possible. We used this configuration to validate the vdW interaction.
The second configuration placed the OH-OH groups close to each other to test hydrogen bonding.
The energy profiles as a function of intermolecular distance are shown in \autoref{fig:nrj_configs}A for the chain-chain configuration and in \autoref{fig:nrj_configs}B for the OH-OH configuration.
We observed good agreement between the MLP and DFT results for the chain-chain interactions, with the minimum at the correct position, albeit a bit deeper than the DFT energies. This could explain the small differences in the heat of vaporization.
On the other hand, the MLP did not reproduce the hydrogen bonding interaction as well as the chain-chain interaction: the minimum was at a shorter distance, and the slope at larger distances was too steep. However, the depth of the minimum seemed to be captured well. Furthermore, the OH-OH interaction could not be described well by the MLP beyond 3~\AA, as this was the short-range cutoff of the MLP. We believe that the steepness in slope comes from the fact that the MLP has to achieve the same depth of the minimum at a shorter distance, leading to a steeper slope.
In addition, molecular bimers were not in the training set, and we believe that the MLP would perform better if trained on a larger set of configurations including dimers. Nevertheless, the MLP was able to capture the main features of the interaction, which is supported by the good performance computing the heat of vaporization.

\FloatBarrier

\onecolumn
\FloatBarrier
\subsection{Errors on energy and forces of MLP}
\begin{figure}[h!]
    \centering
    \includegraphics[width=\hsize]{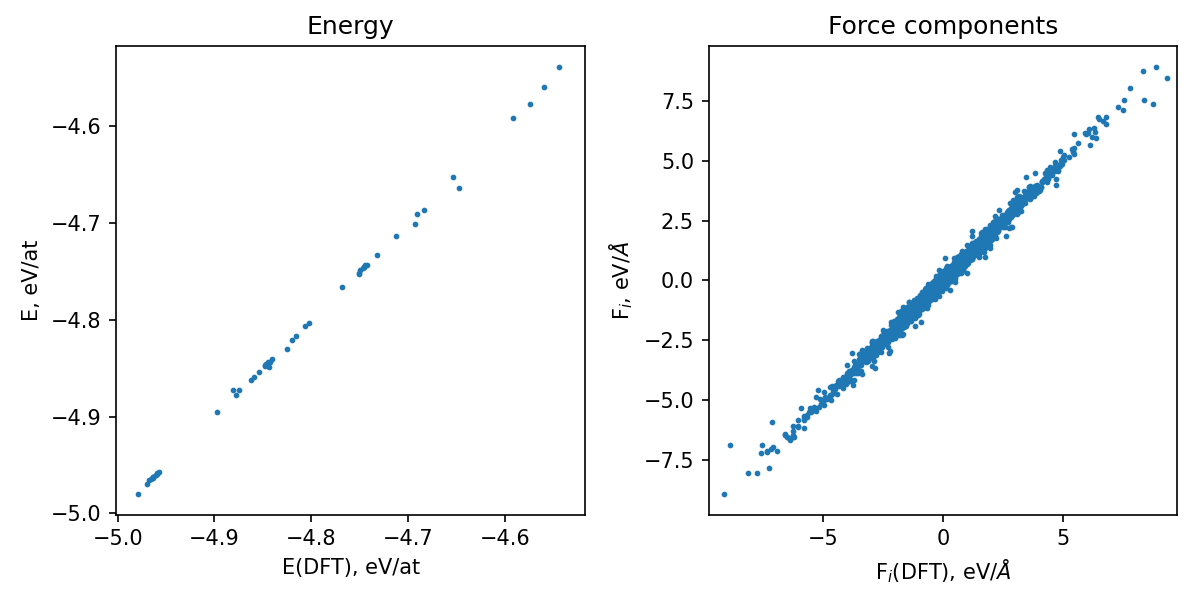}
    \caption{(Left) The energies of test configurations as calculated in DFT against the energies for the same configurations by the best ACE-MLP.
    (Right) Forces in DFT against forces of the ACE-MLP. }
    \label{fig:test_EF-pairplots-best}
\end{figure}

\begin{figure}[h!]
    \centering
    \includegraphics[width=\hsize]{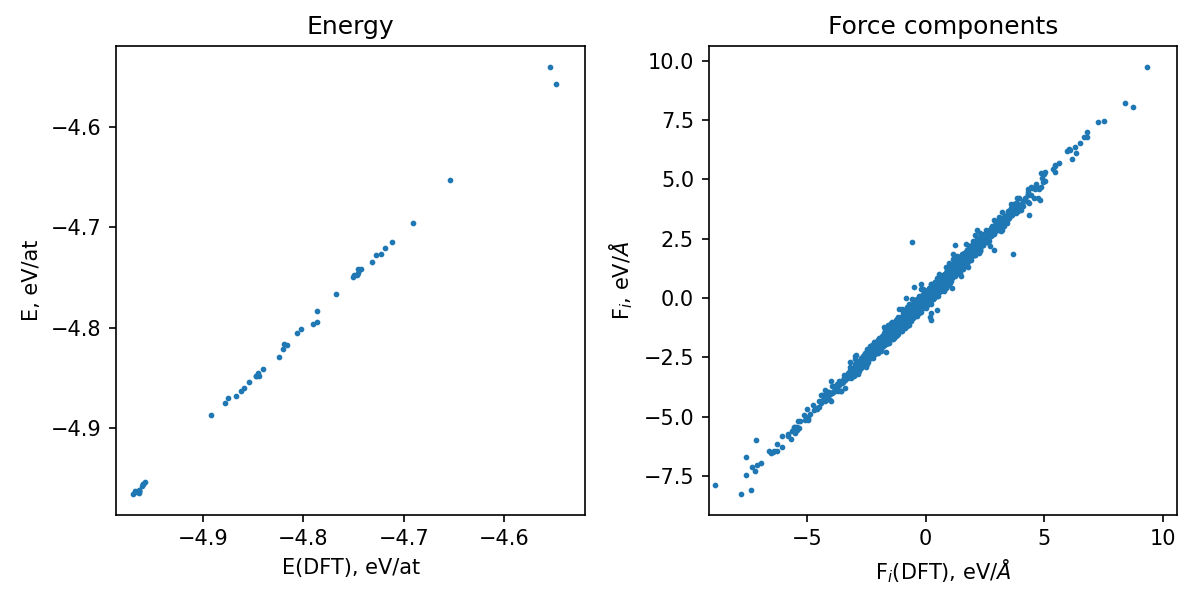}
    \caption{The energies (left) and forces (right) of test configurations as calculated in DFT against the energies for the same configurations by the ACE-MLP with $r_c^s = r_c^l = 3$~\AA.}
    \label{fig:test_EF-pairplots-no-vdw-rc-3}
\end{figure}

\begin{figure}[h!]
    \centering
    \includegraphics[width=\hsize]{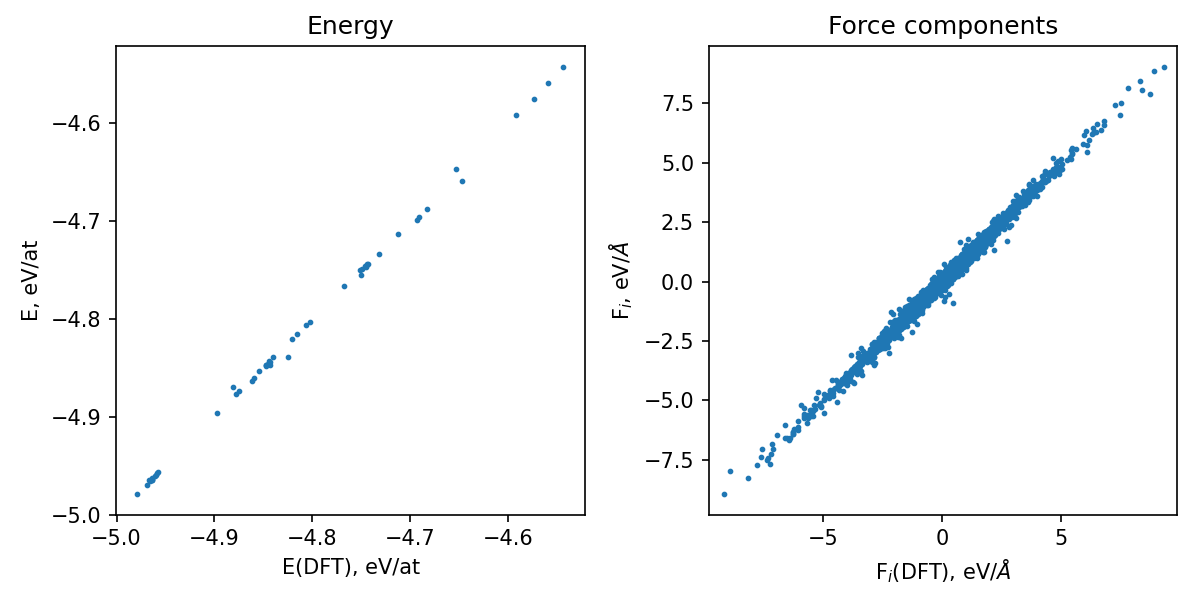}
    \caption{The energies (left) and forces (right) of test configurations as calculated in DFT against the energies for the same configurations by the ACE-MLP with $r_c^s = r_c^l = 5$~\AA.
    Note that this MLP was highly unstable in MD simulations, despite the excellent reproduction of energies and forces.
     \vspace{0.9cm}
     }
    \label{fig:test_EF-pairplots-no-vdw-rc-5}
\end{figure}

\begin{figure}[h!]
    \centering
    \includegraphics[width=\hsize]{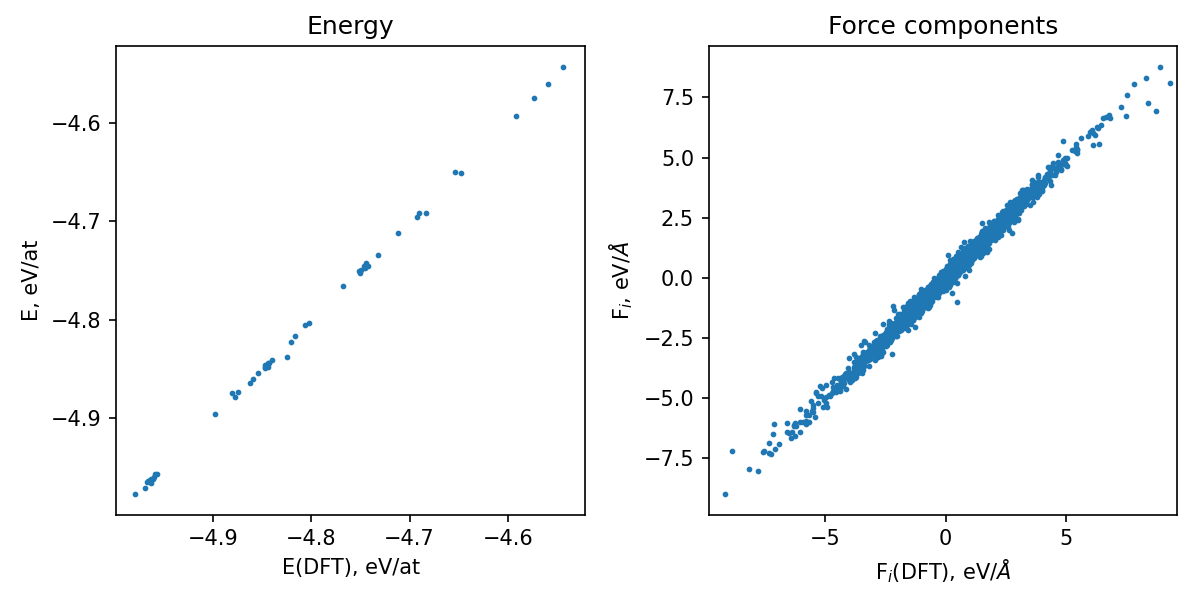}
    \caption{The energies (left) and forces (right) of test configurations as calculated in DFT against the energies for the same configurations by the ACE-MLP with $r_c^s = r_c^l = 7$~\AA.
    Note that this MLP was highly unstable in MD simulations, despite the excellent reproduction of energies and forces.
    \vspace{0.9cm} 
    }
    \label{fig:test_EF-pairplots-no-vdw-rc-7}
\end{figure}

\end{document}